\documentclass[twocolumn,aps,showpacs,prb,tightenlines,amsmath,amssymb]{revtex4}
\usepackage{graphicx}
\usepackage{amssymb}
\usepackage{dcolumn}
\usepackage{amsmath}
\usepackage{bm}
\usepackage{colordvi}
\usepackage{mathrsfs}
\makeatletter

\newcommand{\Rmnum}[1]{\expandafter\@slowromancap\romannumeral #1@}
\makeatother

\begin{document}

\title{Gapped superconductivity with all symmetries in InSb (110) quantum wells
  in proximity to $s$-wave superconductor in Fulde-Ferrell-Larkin-Ovchinnikov
  phase or with a supercurrent}  

\author{F. Yang}
\affiliation{Hefei National Laboratory for Physical Sciences at
Microscale, Department of Physics, and CAS Key Laboratory of Strongly-Coupled
Quantum Matter Physics, University of Science and Technology of China, Hefei,
Anhui, 230026, China}

\author{M. W. Wu}
\thanks{Author to whom correspondence should be addressed}
\email{mwwu@ustc.edu.cn.}

\affiliation{Hefei National Laboratory for Physical Sciences at
Microscale, Department of Physics, and CAS Key Laboratory of Strongly-Coupled
Quantum Matter Physics, University of Science and Technology of China, Hefei,
Anhui, 230026, China}

\date{\today}

\begin{abstract} 
We show that all the singlet even-frequency, singlet odd-frequency, triplet
even-frequency and triplet odd-frequency pairings, and together with the
corresponding order 
parameters (gaps) can be realized in InSb (110) spin-orbit-coupled quantum well
in proximity to
$s$-wave superconductor in
  Fulde-Ferrell-Larkin-Ovchinnikov phase or with a supercurrent. It is
  revealed that with the singlet even-frequency order parameter induced by the proximity effect,
triplet even-frequency pairing is induced due to the broken spin-rotational
symmetry by the spin-orbit coupling.  Since the translational symmetry is
broken by the center-of-mass momentum of Cooper pair in
  Fulde-Ferrell-Larkin-Ovchinnikov phase or with a supercurrent, the singlet odd-frequency
pairing can be induced. With the translational and spin-rotational asymmetries,
the triplet odd-frequency pairing is also realized. Then, we show that the
corresponding order parameters can be obtained from the self-energy of the
electron-electron Coulomb interaction with the dynamic screening. The
singlet and the induced triplet even-frequency order parameters are found to exhibit the
conventional $s$-wave and $p$-wave characters in the momentum space,
respectively. Whereas for the induced odd-frequency order parameters in quantum
well, the singlet and triplet ones show the $p$-wave and $d$-wave
characters, respectively. Moreover, the $p$-wave character of the singlet
odd-frequency order parameter exhibits anisotropy with respect to the
direction of the center-of-mass momentum. While for the triplet one, we find
that $d_{x^2}$-wave and $d_{xy}$-wave characters can be obtained with respect
to the direction of the center-of-mass momentum. We show that at proper density,
the singlet even-frequency order parameter is suppressed and 
the induced singlet odd-frequency, triplet even-frequency and triplet
odd-frequency ones can be detected experimentally.  

\end{abstract}
\maketitle 

\section{Introduction}
In the field of superconductivity, symmetry of the Cooper-pair
wavefunction in spin-, time-, and orbital-spaces has attracted much attention for
the past few decades.  Within the framework of superconductivity theory developed by Bardeen,
Cooper and Schrieffer (BCS),\cite{BCS} it is established that a Cooper pair is
formed by two electrons with momentums ${\bf q+k}$ and ${\bf q-k}$ near the
Fermi surface. Due to the Fermi-Dirac statistics, the
Cooper-pair wavefunction must have sign change in the exchange of the two
electrons. In the spin space, Cooper pair can be classified into either singlet or
triplet type. As for the orbital part of the pair wavefunction, when we
focus on the symmetry with respect to 
the exchange of two momentums ${\bf q+k}$ and ${\bf q-k}$, i.e., ${\bf k}\rightarrow-{\bf k}$,
one can define the parity of the Cooper pair. For the conventional BCS
superconductors like Al, Pb and Nb, in the presence of the translational symmetry (${\bf
  q}=0$) and together with the space-, time-inversion
and spin-rotational symmetries,
the Cooper pairs are in the singlet even-frequency 
even-parity state, in consistent with the Fermi-Dirac statistics. Here, even/odd
frequency refers to the situation that the pair wavefunction is even/odd with respect to the
exchange of time coordinates.  

In 1974, Berezinskii considered the possibility of the triplet even-frequency
pairing with even parity in the observed phase of $^3$He.\cite{TO} After  
that, the possibilities of the Cooper
pair with other symmetries are extensively studied, and from the symmetry
analysis,\cite{Symmetry} Cooper pairs can be classified 
into (i) 
singlet even-frequency (SE) with even parity; (ii) singlet odd-frequency (SO)
with odd parity; (iii) triplet
odd-frequency (TO) with even parity; (iv) triplet even-frequency (TE)
with odd parity.
Specifically, after the proposal by Berezinskii, TO pairing has
been discussed in a wide variety of theoretical
models with spin-rotational and time-inversion
asymmetries,\cite{TO-1,TO-2,TO-3,TO-4,TO-5,TO-6,TO-7,TO-8} e.g., Kondo    
model,\cite{TO-1,TO-3} Hubbard model\cite{TO-2,TO-4,TO-6,TO-7} and heavy Fermion
system.\cite{TO-5,TO-8} Meanwhile, several proposals about the SO pairing have
also been reported in the inhomogeneous systems with space- and time-inversion asymmetries by
introducing effective $p$-wave electron-electron (e-e)
interaction.\cite{SO-1,SO-2,SO-3,SO-4,SO-5} In the presence of the
odd-frequency pairings, by considering the retardation effect of the electron
interaction, odd-frequency gaps or odd-frequency order parameters are
theoretically suggested.\cite{TO-8,SO-2,SO-4,Odd} However, up till now, odd-frequency bulk
superconductor has not yet been realized experimentally. Moreover, it is now
commonly believed that the pairing in superfluid $^3$He is in the TE
type.\cite{He-1,He-2,He-3,He-4,He-5,He-6,He-7,He-8} Recently, much effort has 
been focused on the superconductivity in material Sr$_2$RuO$_4$ due to the
generally recognized similarity to that in
$^3$He.\cite{Sr-1,Sr-2,Sr-3,Sr-4,Sr-5,Sr-6,Sr-7,Sr-8,Sr-9,Sr-10,Sr-11} The 
pairing and order parameter in superconducting Sr$_2$RuO$_4$ are theoretically
suggested\cite{Sr-1,Sr-2,Sr-3,Sr-5} and primarily 
confirmed from recent experiments\cite{Sr-4,Sr-6,Sr-7,Sr-8,Sr-9,Sr-10,Sr-11} to be
the $p$-wave TE type. Furthermore, it is reported very recently that the TE
pairing and order parameter can also be realized in the 
non-centrosymmetric superconductor\cite{TE-1,TE-2,TE-3} with spin-orbit coupling
(SOC)  
existing in nature,\cite{SOC-1,SOC-2,SOC-3} whose experimental confirmations are
still in progress.

Although it is not easy so far to realize odd-frequency superconductivity and/or
triplet one in the uniform bulk system, it is more 
promising to induce these pairings in the inhomogeneous
systems with lower symmetry. Specifically, in the last decade,
the proximity effect has been studied intensively in multilayered
structures consisting of superconductors and non-superconducting systems and it is well
known that the superconducting correlation can penetrate into the normal
region. In superconductor-ferromagnet structure, with the time-inversion
and spin-rotational asymmetries, it is well established
that the TO pairing is induced in
ferromagnet.\cite{ML-1,ML-2,ML-3,ML-4,ML-5,ML-6,ML-7,ML-8,ML-9}  Moreover, it was
predicted that with the inhomogeneous ferromagnet, the induced TO pairing 
can diffuse into the ferromagnet with the longer diffusion length than that of
the SE one.\cite{ML-1} Nevertheless, with the
conventional
$s$-wave electron-electron (e-e) interaction, the TO gap (i.e., the TO order parameter) 
is zero. Similar to the magnetization, the SOC can also break the spin-rotational
symmetry. Together, with the broken space-inversion
symmetry by the SOC, the TE pairing is expected to be
induced,\cite{MLS-0,MLS-1,MLS-2,MLS-3,MLS-4,MLS-5,MLS-6}
which was 
first pointed out by Gor'kov and Rashba in $s$-wave superconductor with the
SOC induced by the absorption of ion.\cite{MLS-0} Then, a great deal of efforts have been
devoted to the multilayered structures consisting of superconductors and
spin-orbit-coupled non-superconducting systems as a natural
extension.\cite{MLS-1,MLS-2,MLS-3,MLS-4,MLS-5,MLS-6} The induced TE pairing is
further proved to possess 
parallel spin projection to the 
effective magnetic field due to the
SOC.\cite{MLS-0,MLS-2,MLS-3,MLS-4,MLS-5,MLS-6} However, even in the
presence of the TE pairing, with the 
momentum-independent $s$-wave e-e interaction, no TE gap (TE order parameter)
is realized.  Nevertheless, de Gennes pointed out that in the
non-superconducting material proximity to superconductor, the pairing
penetrating from superconductor experiences the many-body interaction,\cite{Re}
and hence order parameter can be induced even with a repulsive effective e-e
interaction. Following the work by de Gennes,\cite{Re} it is reported by Yu and
Wu very recently that the TE order parameter is induced in the spin-orbit
coupled quantum well (QW) in proximity to $s$-wave
superconductor.\cite{MLS-6} Specifically, with the induced TE pairing in QW by
the SOC, they showed that from the self-energy of the e-e Coulomb interaction,
the TE order parameter can be induced. 

Except for the multilayered structures, the study of the mixed types of Cooper
pairs in superconductors with vortex structure, where the translational
symmetry is broken, has a long history.\cite{v-1,v-2,v-3,v-4,v-5,v-6,v-7,v-8,v-9,v-10}
It is known that due to the translational asymmetry, the SO pairing is induced
near the vortex core.\cite{Symmetry,v-1,v-2,v-3,v-4,v-5,v-6,v-7,v-8,v-9,v-10}
Besides the vortex structure, the translational symmetry can also be
broken by a supercurrent in superconductors, which leads to the center-of-mass (CM)
momentum of the Cooper pair, as revealed both experimentally\cite{Exper} and
theoretically\cite{Yu} in recent works. Moreover, except for the
extrinsic breakdown of the translational symmetry above, there exists a
high-magnetic-field phase in the  
superconductor, referred as Fulde-Ferrell-Larkin-Ovchinnikov (FFLO) phase,\cite{FF,LO} where
the translational symmetry is spontaneously broken by inducing 
the CM momentum  of the Cooper pair. In the FFLO phase, 
the Zeeman energy leads to different Fermi surfaces for spin-up and -down
electrons, and then by inducing a finite CM momentum of the Cooper pair, the
pairing region between the spin-up and -down electrons near the Fermi surfaces
can be maximized, leading to the free energy minimized. Consequently, there
exist pairing electrons and unpairing ones in the FFLO
phase. Furthermore, with translational and spin-rotational asymmetries by
the magnetic field, all four types of pairings are expected to be 
induced.\cite{FP} Nevertheless, with the
conventional symmetric $s$-wave e-e interaction, only the
SE order parameter exists in the FFLO phase. It is natural to consider the
possibility to induce and manipulate all four types of order parameters, which
may lead to rich physics especially for the quasiparticles.   
Multilayered structures consisting of non-superconducting
systems and $s$-wave superconductors in
FFLO phase or with a supercurrent hence naturally come to our attention. 

In this work, we show that the order parameters containing
all four types of symmetry, i.e., the SE, SO, TE and TO, can be
realized in the two-dimensional electron gas (2DEG) of the spin-orbit-coupled
InSb (110) QW\cite{QW-1,QW-2,QW-3,QW-4,QW-5} in proximity to $s$-wave
superconductors in
FFLO phase or with a supercurrent. Specifically, the SE order parameter can be induced in QW
through the proximity effect. We show that there exist unpairing
regions in the momentum space, where the proximity-induced SE order parameter
vanishes. It is further revealed that the
unpairing regions arise from the FFLO-phase-like blocking in QW. With this
proximity-induced SE order parameter, SO (TE) pairing is induced due to
the broken translational (spin-rotational) symmetry by the 
CM momentum of Cooper pair (SOC).
With the translational and spin-rotational asymmetries, the TO pairing can also
be induced. Then, we
show that from the self-energy due to the e-e Coulomb interaction with the
dynamical screening,\cite{mt1,mt2,mt3} the corresponding order parameters can be induced and the
proximity-induced SE order parameter is also renormalized. Particularly, we reveal
that the odd-frequency order parameters are induced due to the retardation
effect of the Coulomb interaction from the dynamic screening in 2DEG, where the
plasmon effect is important. In addition, the induced triplet order parameters are shown to
possess parallel spin projections to the effective magnetic field due to the
SOC, similar to the previous works.\cite{MLS-0,MLS-2,MLS-3,MLS-4,MLS-5,MLS-6} 

Differing from the
vanishing proximity-induced SE order parameter in the unpairing
regions, it is found that through
the renormalization due to the e-e Coulomb interaction, the SE, SO, TE and TO
ones in QW all have small strengths in the
unpairing regions. Moreover, in the pairing regions, rich behaviors of
the order parameters containing all four types of symmetry are
revealed. Specifically, 
the SE order parameter $\Delta_{\rm SE}$ exhibits an $s$-wave behavior in the momentum space,
while the induced SO (TE) one $\Delta_{\rm SO}$ ($\Delta_{\rm TE}$) shows a $p$-wave
character. Particularly, with the broken translational symmetry by the CM
momentum of Cooper pair, the $p$-wave 
character of the induced SO order parameter shows anisotropy
with respect to the direction of the CM momentum. This is very different from
the TE one, which is determined by the SOC and hence is independent on the direction of CM momentum.
As for the
induced TO order parameter $\Delta_{\rm TO}$, the unconventional $d$-wave character in the
momentum space is revealed and particularly, when the CM momentum is along the $[1{\bar 1}0]$
and $[001]$ directions, the $d_{x^2}$-wave and $d_{xy}$-wave characters 
can be obtained, respectively. This anisotropy of the TO order parameter is further proved to
arise from the unique SOC structure in InSb (110) QW. The specific behaviors of the
order parameters containing all four types of symmetry are summarized in
Table~\ref{Torder}. Furthermore,  
we show that at proper density, the SE
order parameter can be efficiently suppressed through the renormalization from the
repulsive e-e Coulomb interaction, as revealed in the previous work,\cite{MLS-6}
and the induced SO, TE and TO order parameters can be detected and
distinguished experimentally.

This paper is organized as follows. In Sec.~\ref{model}, we introduce our model and lay
out Hamiltonian. In Sec.~\ref{analytic}, we present the analytical results including
the SE, SO, TE and TO pairing functions and the calculation of the self-energy
due to the e-e Coulomb interaction.  The specific numerical results in InSb
(110) QW and analytic analysis are presented in Sec.~\ref{numerical}. We
summarize in Sec.~\ref{summary}. 
    
\begin{table}[htb]
  \caption{Order parameters behaviors in the momentum space. $\Delta^T_{\rm SE}$
    stands for the proximity-induced SE order parameter. $C_{\rm SE}(k,\omega)$,
    $C_{\rm SO}(k,\omega)$, $C_{\rm TE}(k,\omega)$, $C^0_{\rm TO}(k,\omega)$ and $C^1_{\rm TO}(k,\omega)$ are independent
    on the orientation of the momentum.} 
  \label{Torder} 
  \begin{tabular}{l l}
    \hline
    \hline
    Order parameter&\quad~Behavior in the momentum space\\
      \hline
    ${\Delta_{\rm SE}}$&\quad$\Delta^T_{\rm SE}+C_{\rm SE}(k,\omega)$\\ 
    ${\Delta_{\rm SO}}$&\quad$C_{\rm SO}(k,\omega){\bf k{\cdot{q}}}$\\
    ${\Delta_{\rm TE}}$&\quad$C_{\rm TE}(k,\omega){\bf k{\cdot{e_x}}}$\\ 
    ${\Delta_{\rm TO}}$&\quad${C^0_{\rm TO}(k,\omega){\bf q{\cdot}e_x}-C^1_{\rm TO}(k,\omega)({\bf
      k{\cdot}e_x})({\bf k{\cdot}q})}$\\ 
    \hline
    \hline
  \end{tabular}\\
\end{table}

\section{MODEL AND HAMILTONIAN}
\label{model}

In this section, we present the Hamiltonian of the QW in proximity to $s$-wave
superconductors in
FFLO phase or with a supercurrent 
in the Nambu$\otimes$spin space, including the QW ${\hat H_{\rm QW}}$
with the growth direction along the ${\hat z}$-axis, the 
superconductor ${\hat H_{\rm S}}$ in
FFLO phase or with a supercurrent (the CM momentum of
Cooper pair ${\bf q}$ is chosen to be along the in-plane
direction), and the 
tunneling between the QW and superconductor ${\hat H_{\rm T}}$. 
By defining the Nambu spinors in QW {\small {${\hat \Psi}_{\bf
  k}(t)=[\psi_{\uparrow{\bf k+q}}(t),\psi_{\downarrow{\bf 
  k+q}}(t),\psi^{\dagger}_{\uparrow{\bf -k+q}}(t),\psi^{\dagger}_{\downarrow{\bf
  -k+q}}(t)]^{T}$}} and in superconductor {\small{${\hat
\Phi}_{\bf 
  p}(t)=[\phi_{\uparrow{\bf p+q}}(t),\phi_{\downarrow{\bf 
  p+q}}(t),\phi^{\dagger}_{\uparrow{\bf -p+q}}(t),\phi^{\dagger}_{\downarrow{\bf
  -p+q}}(t)]^{T}$}}, with ${\bf 
k}=k_x{\bf e_x}+k_y{\bf e_y}$
and ${\bf p}=p_x{\bf e_x}+p_y{\bf e_y}+p_z{\bf e_z}$
being the momentums of the electrons in QW and superconductor, respectively, these
Hamiltonians are given in the following. 

The Hamiltonian of the spin-orbit-coupled QW is given by\cite{MLS-6}
\begin{equation}
{\hat H_{\rm QW}}={\hat H^{\rm k}_{\rm QW}}+{\hat H^{\rm SOC}_{\rm QW}}+{\hat H^{\rm ee}_{\rm QW}},
\end{equation}
where ${\hat H^{\rm k}_{\rm QW}}$, ${\hat H^{\rm SOC}_{\rm QW}}$, ${\hat H^{\rm
    ee}_{\rm QW}}$ are the kinetic energy, the SOC, and the e-e Coulomb interaction,
respectively:
\begin{eqnarray}
{\hat H^{\rm k}_{\rm QW}}&=&\frac{1}{2}\int{d{\bf k}}{\hat \Psi}^{\dagger}_{\bf
  k}(t)\left(\begin{array}{cc}
\xi^c_{\bf k+q} & 0\\
0 & \xi^c_{\bf -k+q}\\
\end{array}\right)\rho_3{\hat \Psi}_{\bf k}(t),\\
{\hat H^{\rm SOC}_{\rm QW}}&=&\frac{1}{2}\int{d{\bf k}}{\hat \Psi}^{\dagger}_{\bf
  k}(t)\left(\begin{array}{cc}
h_{\bf k+q} & 0\\
0 & h_{\bf -k+q}\\
\end{array}\right)\rho_3{\hat \Psi}_{\bf k}(t),\\
{\hat H^{\rm ee}_{\rm QW}}&=&\frac{1}{8}\int{{d{\bf k}}{d{\bf k'}d{\bf q'}}}{V_{\bf q'}(t-t')}[{\hat \Psi}^{\dagger}_{\bf
  k+q'}(t)\rho_3{\hat \Psi}_{\bf k}(t)]\nonumber\\
  &&\mbox{}\times[{\hat \Psi}^{\dagger}_{\bf
  k'-q'}(t')\rho_3{\hat \Psi}_{\bf k'}(t')].
\end{eqnarray}
Here, $\xi^c_{\bf k}=\varepsilon^c_{\bf k}-\mu_c$ and $\varepsilon^c_{\bf
  k}=k^2/(2m^*_c)$ with $m^*_c$ and $\mu_c$ being the 
effective mass and chemical potential 
of the electron in QW, respectively; $h_{\bf k}$ represents the SOC;
$\rho_3=\sigma_0\otimes\tau_3$; 
$\sigma_i$ and $\tau_i$ stand for the Pauli matrices in spin and particle-hole
spaces, respectively. The e-e Coulomb interaction with the dynamic screening
considered is given by $V_{\bf 
  k}(t-t')=\int{d\omega}e^{-i\omega(t-t')}V_{\bf k}(\omega)$ with 
\begin{equation}
V_{\bf k}(\omega)=\frac{V^0_{\bf k}}{\epsilon({\bf k},\omega)}.
\end{equation}
Here, $V^0_{\bf k}={2\pi{e^2}}/({\epsilon_0\kappa_0q})$ stands for the
unscreened Coulomb potential in 2D system; $\epsilon_0$ and $\kappa_0$
represent the vacuum permittivity and relative 
dielectric constant, respectively; $\epsilon({\bf q},\omega)$ is
the dynamic dielectric function. In the long-wavelength
limit ($\omega>qv_F$), based on the linear response theory,\cite{mt1,mt2,mt3}
the expression of 
$V_{\bf k}(\omega)$ can be given by 
\begin{equation}
\label{Coulomb}
V_{\bf k}(\omega)=V^{\rm re}_{\bf k}+V^{\rm at}_{\bf k}(\omega)
\end{equation}
where $V^{\rm re}_{\bf k}=V^0_{\bf k}$; 
$V^{\rm at}_{\bf k}=V^0_{\bf k}|\omega^{pl}_{\bf
    k}|^2/(\omega^2-|\omega^{pl}_{\bf
    k}|^2)$, which has been revealed in the previous work\cite{at} to act as a
  retarded attractive potential; $\omega^{pl}_{\bf
  k}=\sqrt{2\pi{e^2}nk/(m^*_c\epsilon_0\kappa_0)}$ is the plasma frequency with
  $n$ being the density of the electrons. 

The Hamiltonian of the $s$-wave
superconductor in
FFLO phase or with a supercurrent is expressed as\cite{FF,LO}   
\begin{eqnarray}
&&{\hat H_{S}}=\int\frac{d{\bf p}}{2}\nonumber\\
&&\mbox{}\times{\hat \Phi}^{\dagger}_{\bf
  p}(t)\left(\begin{array}{cc}
\xi^s_{\bf p+q}+h_B\sigma_z & \Delta_0i\sigma_2\\
\Delta^*_0i\sigma_2 & \xi^s_{\bf -p+q}+h_B\sigma_z\\
\end{array}\right)\rho_3{\hat \Phi}_{\bf p}(t),~~~~~
\end{eqnarray}
where $\xi^s_{\bf p}=\varepsilon^s_{\bf p}-\mu_s$ and $\varepsilon^s_{\bf
  p}=p^2/(2m^*_c)$ with $m^*_s$ and $\mu_s$ representing the
effective mass and chemical potential of the electrons in superconductor,
respectively; $\Delta_0$ denotes the singlet gap and
 $h_B$ stands for the Zeeman energy in the
superconductor. In this work, we mainly consider
the physics of the translational asymmetry by the CM momentum of Cooper
pair, which can be induced by either the FFLO phase or a supercurrent in
superconductors, as mentioned in the introduction. The magnetic field in QWs, which
can lead to TO pairing in QWs as
revealed in the previous 
works,\cite{ML-1,ML-2,ML-3,ML-4,ML-5,ML-6,ML-7,ML-8,ML-9} is not included
here. Particularly, for the case of superconductors in FFLO phase (or with a supercurrent), our
proposal can be realized in N/S/F (or N/S) structures, hence the magnetic
field in QWs is not considered.

The tunneling Hamiltonian between the QW and
superconductor reads\cite{MLS-6,sa1,sa2,sa3}
\begin{eqnarray}
&&{\hat H_T}=\int{d{\bf k}dp_z}\nonumber\\
&&\mbox{}\times\left[\Psi^{\dagger}_{\bf k}(t)t_0\rho_3\Phi_{({\bf k},p_z)}(t)+\Phi^{\dagger}_{({\bf k},p_z)}(t)t_0^*\rho_3\Psi_{\bf k}(t)\right],\label{tunneling}
\end{eqnarray}
with $t_0$ being the tunneling matrix element.

\section{Analytical Results}
\label{analytic}

We first start our investigation from the effective BdG
Hamiltonian in InSb (110) QW following the approach in the previous
work\cite{MLS-6} by using the 
equilibrium Green functions,\cite{mt1,mt2,mt3} and show that the singlet and triplet order
parameters can be induced from the self-energy due to the e-e Coulomb
interaction. In InSb (110) QW, the SOC $h_{\bf
  k}=\gamma_Dk_x(k_x^2-2k_y^2-\langle{k_z^2}\rangle)\sigma_z$,
with $\gamma_D$ being the Dresselhaus coefficient.\cite{SOC}  Particularly, since
$\langle{k^2_z}\rangle\gg{k^2_F}$ in QW,\cite{QW-1,QW-2,QW-3,QW-4}
one approximately has   
\begin{equation}
h_{\bf k}=-\gamma_D\langle{k^2_z}{\rangle}k_x\sigma_z.\label{hk}
\end{equation}
Here, the in-plane coordinate axes
are set as $x\parallel[1{\bar 1}0]$ 
and $y\parallel[001]$ with ${\bf e_x}$ (${\bf e_y}$) being the unit vector along the ${\hat
  x}$ (${\hat y}$) direction. Furthermore, it has been shown in the
    previous work\cite{MLS-6}
that in InSb QWs, the self-energy due to the electron-phonon interaction is much smaller
than that due to the e-e Coulomb interaction at low temperature, and hence only
the Coulomb interaction needs to be considered in this work.

\subsubsection{Effective BdG Hamiltonian in QW}
\label{SPEC}
In the Nambu$\otimes$spin space, the equilibrium Green function in the momentum
space is given by 
\begin{equation}
G_{\bf k}(t)=-i\rho_3\langle{T{\hat \Psi}_{\bf
  k}(t){\hat \Psi}^{\dagger}_{\bf
  k}(0)}\rangle, 
\end{equation}
where $T$ represents the time-ordering operator; $\langle~\rangle$
denotes the
ensemble average. By expressing 
\begin{equation}
G_{\bf k}(t)=\left(\begin{array}{cc}
g_{\bf k}(t)& f_{\bf k}(t)\\ 
f^{\dagger}_{\bf -k}(t)& {g^{\dagger}}_{\bf -k}(t)\\ 
\end{array}\right),
\end{equation} 
one can obtain the normal Green function $g_{\bf k}(t)$ and anomalous Green
function $f_{\bf k}(t)$.\cite{MLS-0,MLS-6,mt1,mt2,mt3}

In the frequency space $G_{\bf
  k}(\omega)=\int^{\infty}_{-\infty}e^{i\omega{t}}G_{\bf k}(t)$, the Gor'kov
equation\cite{GE} in QW is given by 
\begin{equation}
\label{frequency}
(\omega\rho_3-H^{\rm k}_{\rm QW}-H^{\rm
  SOC}_{\rm QW})G^0_{\bf k}(\omega)=1,
\end{equation}
with $G^0_{\bf k}(\omega)$ being the free Green function. When the interaction is
considered, from the Dyson equation $G_{\bf k}(\omega)=G^0_{\bf
  k}(\omega)+G^0_{\bf k}(\omega){\hat \Sigma}({\bf k},\omega)G_{\bf
  k}(\omega)$,\cite{mt1,mt2,mt3} one obtains 
\begin{equation}
\label{Gorkov}
\left[\omega\rho_3-H^{\rm k}_{\rm QW}-H^{\rm
  SOC}_{\rm QW}-{\hat \Sigma}({\bf k},\omega)\right]G_{\bf k}(\omega)=1,
\end{equation}
where ${\hat \Sigma}({\bf k},\omega)$ are the self-energies due to $H_T$ and
$H^{\rm ee}_{\rm QW}$.
By comparing Eq.~(\ref{Gorkov}) with Eq.~(\ref{frequency}), the
effective BdG Hamiltonian in QW is obtained:\cite{MLS-6} 
\begin{equation}
\label{BdG}
{\hat H^{\rm BdG}_{\rm QW}}={\hat H^0_{\rm QW}}+{\hat \Sigma}({\bf k},\omega),
\end{equation}
from which one can obtain the singlet and triplet order parameters and calculate
the energy-spectra of the elementary excitation.

\subsubsection{Self-energy due to the proximity effect}
\label{sdpe}
Following the previous work,\cite{MLS-6} by approximately considering that the
e-e Coulomb
interaction is weaker than the proximity effect, we first calculate the
self-energy due to the proximity effect without the e-e Coulomb interaction to determine
the Green function, and then obtain the self-energy due to the e-e
Coulomb interaction.

The self-energy due to the tunneling is calculated based on Hamiltonian
Eq.~(\ref{tunneling}), written as
\begin{equation}
\label{SelfT}
{\hat \Sigma_{T}}({\bf k},\omega)=|t_0|^2\int{dp_z}{G^s_{{\bf k}+p_z{\bf {e_z}}}(\omega)}.
\end{equation}
Here, $G^s_{\bf p}(\omega)=\int^{\infty}_{-\infty}e^{i\omega{t}}G^s_{\bf p}(t)$
with $G^s_{\bf p}(t)$ being the Green function in 
superconductor, written as 
\begin{equation}
G^{s}_{\bf p}(t)=-i\rho_3\langle{T{\hat \Phi}_{\bf
  p}(t){\hat \Phi}^{\dagger}_{\bf
  p}(0)}\rangle=\left(\begin{array}{cc}
g^s_{\bf p}(t)& f^s_{\bf p}(t)\\ 
{{f}^{s}}^{\dagger}_{\bf -p}(t)& {{g}^s}^{\dagger}_{\bf -p}(t)\\ 
\end{array}\right).
\end{equation}
$g^s_{\bf p}(t)$ and $f^s_{\bf p}(t)$ are the normal and anomalous Green
functions, respectively.
Specifically, in the frequency space, the anomalous Green
function is given by  
\begin{equation}
f^{s}_{\bf p}(\omega)=\left(\begin{array}{cc}
 0 & \frac{\Delta_0}{D_{{\bf p},\omega}(h_B)}\\
\frac{-\Delta_0}{D_{{\bf p},\omega}(-h_B)}& 0\\
\end{array}\right),
\end{equation}
where $D_{{\bf p},\omega}(\pm{h_B})=(\omega-\xi^s_{\bf
  p+q}\mp{h}_B)(-\omega-\xi^s_{\bf 
  p-q}\pm{h}_B)+|\Delta_0|^2$. 
By neglecting the diagonal terms in Eq.~(\ref{SelfT}) which are marginal at the
weak coupling limit,\cite{sa1,sa2,sa3} the self-energy in QW due to the
proximity effect can be obtained 
\begin{equation}
\label{self-t}
{\hat \Sigma_{T}}({\bf k},\omega)=\left(\begin{array}{cc}
0 & {\hat \Delta^T}({\bf k},\omega)\\
{\hat {\Delta^T}}^*({\bf k},\omega)  &  0\\
\end{array}\right),
\end{equation}
where 
\begin{equation}
\label{amp}
{\hat \Delta^T}({\bf k},\omega)=\int{dp_z}{{|t_0|^2}f^{s}_{{\bf k}+p_z{\bf e_z}}(\omega)}.
\end{equation}
In the weak coupling limit, the order parameters
induced in QW is much smaller than $\Delta_0$.\cite{sa1,sa2,sa3} Moreover,
we focus on the 
low-frequency regime [$\omega{\le}\Delta({\bf k},\omega)$] where the main
physics happens. Consequently, the frequency is much smaller than $\Delta_0$. In
this case, the frequency dependence of $\Sigma_{T}({\bf k},\omega)$ can be
neglected.\cite{MLS-6,sa1,sa2,sa3} Therefore,
by considering the small ${\bf k}$ in QW ($\varepsilon^s_{\bf k}\ll\mu_s$),  
one obtains (refer to Appendix~\ref{AP-de})
\begin{equation}
 \label{ggll}
{\hat \Delta^T({\bf k},0)}={\Delta^T_{\rm SE}({\bf k})}i\sigma_2=\frac{\Delta_0|{\tilde t}|^2}{\varepsilon^s_{\bf k+q}\varepsilon^s_{\bf
  k-q}+|\Delta_0|^2-h_B^2}i\sigma_2,
\end{equation}
acting as the SE order parameter [${\Delta^T_{\rm SE}({\bf k})}$] in QW due to the proximity
effect. Here, ${\tilde t}$ is the effective tunneling matrix
element (given also in Appendix~\ref{AP-de}).

Based on Eqs.~(\ref{Gorkov}) and~(\ref{self-t}), in the absence of the e-e Coulomb
interaction, one can derive the Green
function $G_{\bf k}(\omega)$ in QW with the proximity-induced SE order
parameter included. Particularly, the anomalous Green function is
\begin{equation}
\label{smag}
f_{\bf k}(\omega)=\left(\begin{array}{cc}
 0 & -\frac{\Delta^T_{\rm SE}({\bf k})}{(\omega-E^+_{+{\bf k}})(\omega+E^+_{-{\bf k}})}\\
\frac{\Delta^T_{\rm SE}({\bf k})}{(\omega-E^-_{+{\bf k}})(\omega+E^-_{-{\bf k}})}& 0\\
\end{array}\right),
\end{equation}   
where $E^{\mu}_{\nu{\bf k}}=\sqrt{(\frac{\xi^c_{\bf k+q}+\xi^c_{\bf
      k-q}}{2}+\mu\frac{h_{\bf k+q}+h_{\bf
      k-q}}{2})^2+|\Delta^T_{\rm SE}({\bf k})|^2}+\nu\frac{\xi^c_{\bf k+q}-\xi^c_{\bf
      k-q}}{2}+\mu\nu\frac{h_{\bf k+q}-h_{\bf
      k-q}}{2}$ are the quasi-particle energy spectra ($\mu,\nu=\pm1$) in
  QW. It is noted that there exist regions with $E^{\mu}_{\nu{\bf  
      k}}<0$ in the momentum space, where the quasi-particle energies are below the
  Fermi surface. Here,
  following the FFLO idea,\cite{FF,LO} in such regions, the Cooper pairs must be
  broken since these quasi-particle states are perfectly blocked by the
  electrons. Hence, we consider the regions with
  $E^{\mu}_{\nu{\bf 
      k}}<0$ are the unpairing regions where the tunneling
  is blocked and then the proximity-induced SE order parameter should vanish.

 Consequently, the
  proximity-induced SE order parameter is given by
\begin{equation}
\label{pso}
{\Delta^T_{\rm SE}({\bf k})}=\Delta_0\frac{{\tilde t}^2}{\varepsilon^s_{\bf k+q}\varepsilon^s_{\bf
  k-q}+|\Delta_0|^2-h_B^2}{\hat \delta},
\end{equation}
with $\delta_{\bf k}=\Pi_{\mu=\pm,\nu=\pm}\theta(E^{\mu}_{\nu{\bf k}})$ being the
depairing operator.\cite{FF} Here, $\theta(x)$ is the step function.

From Eqs.~(\ref{self-t}) and~(\ref{ggll}), it is noted that only
the SE order parameter exists. However, it is shown in the following that all
four types of the order parameters can be realized in QW when the 
e-e Coulomb
interaction is considered. 

\subsubsection{Self-energy due to the e-e Coulomb interaction}
 
The self-energy in the frequency space due to the e-e Coulomb interaction reads\cite{mt1,mt2,mt3} 
\begin{equation}
\label{self-ee}
{\hat \Sigma_{\rm ee}}({\bf k},\omega)=\frac{1}{4}\int\frac{d{\bf
    k'}}{(2\pi)^2}\frac{d\omega'}{2\pi}V_{\bf k-k'}(\omega-\omega')G({\bf k'},\omega').
\end{equation}
From Eq.~(\ref{self-ee}), one observes that from the self-energy due to the e-e Coulomb
interaction, the normal Green function and anomalous Green function are both
renormalized. Consequently, the SOC strength, the effective mass, the
zero-energy point and the SE order parameter are renormalized. Moreover, the SO, TO
and TE order parameters can be induced due to the existence of the corresponding
pairing. In this work, we focus on the renormalization-induced order parameters,
and neglect the renormalization of the SOC strength, effective mass and zero-energy point.

From Eqs.~(\ref{smag}) and~(\ref{self-ee}), the renormalization-induced order
parameters can be obtained. Specifically, after the frequency integration, the
renormalization-induced singlet order parameter [$\Delta_s({\bf k},\omega)$] and triplet
one with zero spin projection [$\Delta_t({\bf k},\omega)$] are written as
\begin{eqnarray}
&&\Delta_{s}({\bf k},\omega)=\sum_{\mu=\pm}\int\frac{d{\bf
    k'}}{32\pi^2}{{V^0_{\bf k-k'}}\Delta^T_{\rm SE}({\bf k'})}\nonumber\\
&&\mbox{}\times\bigg[\frac{1}{E^{\mu}_{+{\bf k'}}+E^{\mu}_{-{\bf
      k'}}}\left(\sum_{\nu=\pm}\frac{\omega^{pl}_{\bf k-k'}}{E^{\mu}_{\nu{\bf
      k'}}+\omega^{pl}_{\bf k-k'}-\nu\omega}-1\right)\nonumber\\
&&\mbox{}+\sum_{\nu=\pm}\frac{|E^{\mu}_{\nu{\bf k'}}-\nu\omega|^2}{|E^{\mu}_{\nu{\bf
      k'}}-\nu\omega|^2-|\omega^{pl}_{\bf k-k'}|^2}\frac{f(E^{\mu}_{\nu{\bf
      k'}})}{E^{\mu}_{+{\bf k'}}+E^{\mu}_{-{\bf k'}}}\nonumber\\
&&\mbox{}+\sum_{\nu=\pm}\frac{E^{\mu}_{\nu{\bf k'}}-\nu\omega}{|E^{\mu}_{\nu{\bf
      k'}}-\nu\omega|^2-|\omega^{pl}_{\bf k-k'}|^2}\frac{n(\omega^{pl}_{\bf
    k-k'})\omega^{pl}_{\bf
    k-k'}}{E^{\mu}_{+{\bf k'}}+E^{\mu}_{-{\bf k'}}}\bigg],~~~\label{deltas}\\
&&\Delta_t({\bf k},\omega)=\sum_{\mu=\pm}\mu\int\frac{d{\bf
    k'}}{32\pi^2}{{V^0_{\bf k-k'}}\Delta^T_{\rm SE}({\bf k'})}\nonumber\\
&&\mbox{}\times\bigg[\frac{1}{E^{\mu}_{+{\bf k'}}+E^{\mu}_{-{\bf
      k'}}}\left(\sum_{\nu=\pm}\frac{\omega^{pl}_{\bf k-k'}}{E^{\mu}_{\nu{\bf
      k'}}+\omega^{pl}_{\bf k-k'}-\nu\omega}-1\right)\nonumber\\
&&\mbox{}+\sum_{\nu=\pm}\frac{|E^{\mu}_{\nu{\bf k'}}-\nu\omega|^2}{|E^{\mu}_{\nu{\bf
      k'}}-\nu\omega|^2-|\omega^{pl}_{\bf k-k'}|^2}\frac{f(E^{\mu}_{\nu{\bf
      k'}})}{E^{\mu}_{+{\bf k'}}+E^{\mu}_{-{\bf k'}}}\nonumber\\
&&\mbox{}+\sum_{\nu=\pm}\frac{E^{\mu}_{\nu{\bf k'}}-\nu\omega}{|E^{\mu}_{\nu{\bf
      k'}}-\nu\omega|^2-|\omega^{pl}_{\bf k-k'}|^2}\frac{n(\omega^{pl}_{\bf
    k-k'})\omega^{pl}_{\bf
    k-k'}}{E^{\mu}_{+{\bf k'}}+E^{\mu}_{-{\bf k'}}}\bigg].~~~\label{deltat}
\end{eqnarray}
where $f(x)=[{\rm exp}(\beta{x})+1]^{-1}$ is the Fermi-Dirac distribution
function and
$n(x)=[{\rm exp}(\beta{x})-1]^{-1}$ stands for the Bose-Einstein distribution
function; $\beta=1/(k_BT)$ with $k_B$ representing the Boltzmann constant and
$T$ being the temperature.

\begin{table}[htb]
  \caption{Parameters used in our calculation. Note that $m_0$ stands for the
    free electron mass and $a$ is the well width. } 
  \label{parametertable} 
  \begin{tabular}{l l l l}
    \hline
    \hline
    $m^*_c/m_0$&\quad$0.015^a$&\quad\quad$m_s^*/m_0~$&\quad$3.2^b$\\ 
    $\kappa_0$&\quad$16.0^a$&\quad\quad$\Delta_0~$(meV)&\quad$5^b$\\
    $\gamma_D~$(eV${\cdot}${\AA}$^3$)&\quad$79.4^c$&\quad\quad$h_B/\Delta_0$&\quad$0.6^d$\\
    $a~$(nm)&\quad$2$&\quad\quad$T~$(K)&\quad$0.5$\\
    $|{\tilde t}|^2/\Delta^2_0$&\quad$0.5$&\quad\quad$n_0~$(cm$^{-2}$)&\quad$2.5\times10^{9}$\\
    \hline
    \hline
  \end{tabular}\\
  \quad$^a$ Refs.~\onlinecite{parameter1,parameter2}. \quad$^b$
  Ref.~\onlinecite{Nb}. \quad$^c$ Ref.~\onlinecite{SOC}. \quad$^d$
  Refs.~\onlinecite{FF,LO}.  
\end{table}

\section{NUMERICAL RESULTS}
\label{numerical}

In this section, by calculating Eqs.~(\ref{deltas}) and~(\ref{deltat}) explicitly, 
we investigate the order parameters in 2DEG of the spin-orbit-coupled InSb (110)
QW in proximity to $s$-wave
  superconductor in FFLO 
  phase or with a supercurrent. All the material parameters
used in our calculation are listed in Table~\ref{parametertable}. As mentioned
above, the frequency $\omega$ in the calculation is chosen to be smaller than
$\Delta_0$.

\begin{figure}[htb]
  {\includegraphics[width=8.5cm]{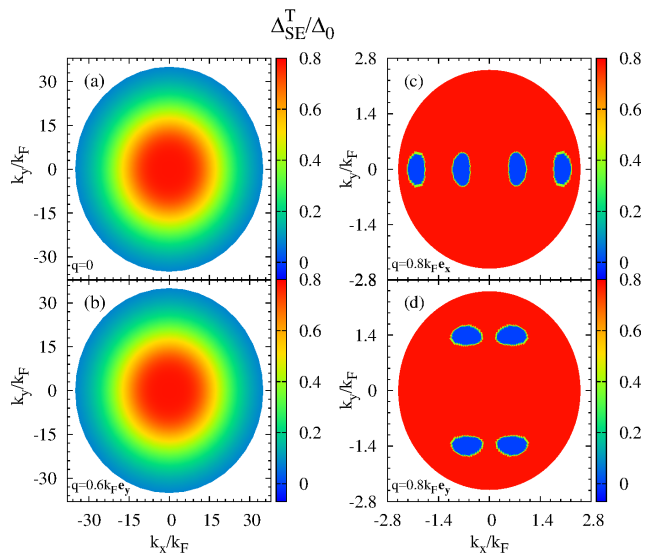}}
\caption{Momentum
dependence of the proximity-induced SE order parameter $\Delta^T_{\rm SE}({\bf
  k})$ at (a) ${\bf q}=0$, (b) ${\bf q}=0.6k_F{{\bf e}_x}$, (c) ${\bf
  q}=0.8k_F{{\bf e}_x}$ and (d) ${\bf q}=0.8k_F{{\bf e}_y}$. $n=5n_0$.}   
\label{figyw1}
\end{figure}

\subsection{SE ORDER PARAMETER}
\label{SE-order}

In this part, we investigate the SE order parameters in QW, including the
proximity-induced and renormalization-induced ones. The proximity-induced
SE order parameter ${\Delta^T_{\rm SE}({\bf k})}$ is obtained from
Eq.~(\ref{pso}) while the renormalization-induced one $\Delta^{R}_{\rm SE}({\bf
  k},\omega)$ is obtained by calculating $\Delta_s({\bf k},\omega)$ 
[Eq.~(\ref{deltas})] explicitly: 
\begin{equation}
  \label{SE}
\Delta^{R}_{\rm SE}({\bf k},\omega)=[{\Delta_{s}({\bf k},\omega)+\Delta_{s}({\bf
    k},-\omega)}]/2.
\end{equation}

\subsubsection{Momentum dependence of the proximity-induced SE order parameter: $s$-wave character}

We first focus on the
proximity-induced SE order parameters, whose momentum
dependences at
different CM momentums ${\bf q}$ are plotted in
Fig.~\ref{figyw1} at $n=5n_0$. We find that when
$q<0.75k_F$,  
the proximity-induced SE order parameters $\Delta^T_{\rm SE}(\bf k)$ show an
$s$-wave character in the momentum space. Moreover, from Eq.~(\ref{pso}),
    with $|\Delta_0|\approx\varepsilon^s_{{16k_F}}$, 
    one finds that the strengths of the proximity-induced SE order parameters are 
marginally influenced when $k\ll15k_F$ but decrease monotonically when $k>15k_F$, as shown in
Figs.~\ref{figyw1}(a) and~(b). In addition, when $k>15k_F$, one has $q\ll{k}$.
Consequently, the dependence of $\Delta^T_{\rm SE}(\bf k)$ on ${\bf q}$ is 
indistinguishable, as shown by the comparison between Figs.~\ref{figyw1}(a)
and~(b).  

Furthermore, it is found that when $q>0.75k_F$, there exist unpairing regions in the 
momentum space, where the proximity-induced SE order parameters vanish, as shown
by the blue regions in Figs.~\ref{figyw1}(c) and~(d) at $q=0.8k_F$. This is
justified by the fact that the positions of the unpairing regions in the
momentum space coincide with those of the regions where the depairing operator
is zero (shown in Appendix~\ref{AP-mo}). As for the pairing region, the
proximity-induced SE order parameter shows the similar behaviors to that at
$q<0.75k_F$.

Consequently, the
proximity-induced SE order parameter in the 
pairing region in QW can be considered as a constant $\Delta^T_{\rm SE}$ when
$k\ll15k_F$. It is further noted that this conclusion is in consistent with 
the constant approximation in the previous works.\cite{MLS-6,sa1,sa2,sa3}

\begin{figure}[htb]
  {\includegraphics[width=8.5cm]{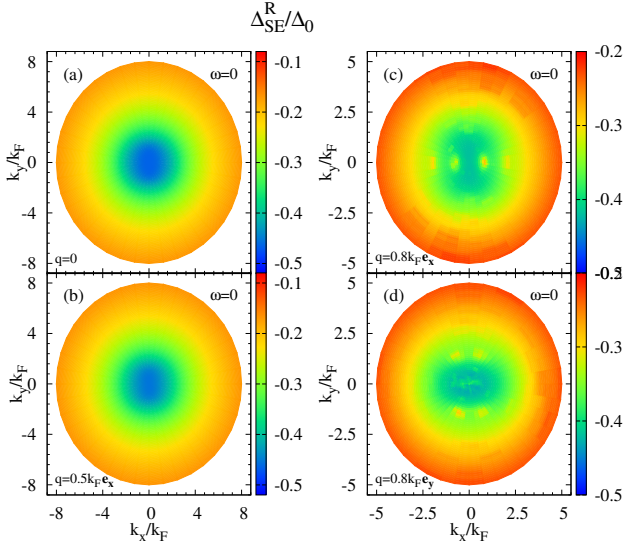}}
\caption{Momentum dependence of the
renormalization-induced SE order parameter $\Delta^R_{\rm SE}({\bf k},\omega=0)$ at (a)
${\bf q}=0$, (b)
${\bf q}=0.5k_F{\bf e_x}$, (c)
${\bf q}=0.8k_F{\bf e_x}$ and (d)
${\bf q}=0.8k_F{\bf e_y}$. $n=5n_0$.}   
\label{figyw2}
\end{figure}

\begin{figure}[htb]
  {\includegraphics[width=9cm]{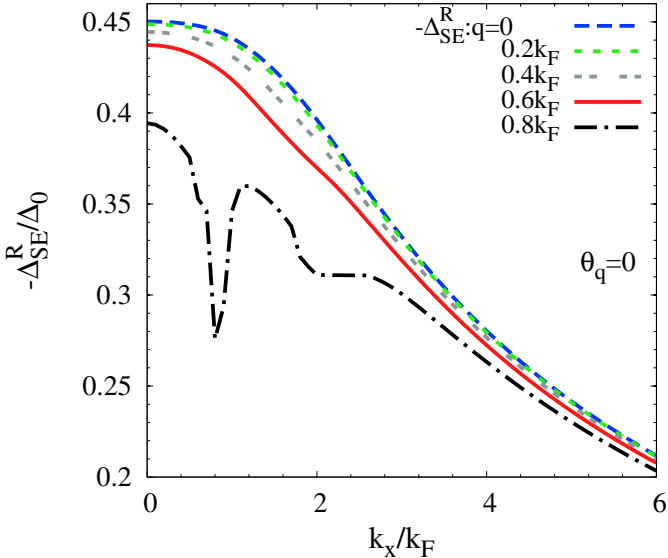}}
\caption{Momentum-magnitude dependence of the strength for the renormalization-induced SE order
  parameter $|\Delta^R_{\rm SE}({\bf
    k}=k{\bf e_x},\omega=0)|=-\Delta^R_{\rm SE}({\bf
    k}=k{\bf e_x},\omega=0)$ at different CM momentums $q{\bf
    e_x}$. $n=5n_0$.}   
\label{figyw3}
\end{figure}

\subsubsection{Momentum dependence of the renormalization-induced SE order 
  parameter: $s$-wave character}

We next discuss the renormalization-induced SE order parameter. The momentum dependences of the
renormalization-induced SE order  
parameter at different CM momentums are
plotted in Fig.~\ref{figyw2} when $n=5n_0$. As shown in Figs.~\ref{figyw2}(a) and~(b), when $q<0.75k_F$, the
renormalization-induced SE order parameters in the momentum space exhibit an
$s$-wave character: $\Delta^R_{\rm SE}=C_{\rm SE}(k,\omega)$ with $C_{\rm
  SE}(k,\omega)$ being independent on the orientation of the
momentum, similar to
the proximity-induced one. Moreover, $\Delta^R_{\rm SE}$ is always
in the opposite sign against the proximity-induced SE order parameter
$\Delta^T_{\rm SE}$ as the renormalization
from the repulsive e-e Coulomb 
interaction. 

When $q>0.75k_F$,
we find that there exist four regions in the momentum space where the renormalization-induced
SE order parameters have smaller strengths than those in the region nearby [shown by the four
yellow regions in the regime $k<2.1k_F$ in Figs.~\ref{figyw2}(c) or (d)]. The
positions of these regions exactly correspond to the unpairing regions where the
proximity-induced SE order parameters vanish [shown by the blue regions in
Figs.~\ref{figyw1}(a) and (b) correspondingly]. This can be understood from the
fact that the 
renormalization of the e-e Coulomb interaction to the SE order parameter leads
to the contribution from the pairing regions to the unpairing ones. 

\subsubsection{Momentum-magnitude dependence of the renormalization-induced SE order 
  parameter}

The momentum-magnitude 
dependences of the strength for the
renormalization-induced SE order  
parameter $-\Delta^R_{\rm
  SE}({\bf k}=k{\bf e_x},\omega)$ are shown in Fig.~\ref{figyw3} at $n=5n_0$. It
is first noted that at
${\bf q}=0.8k_F{\bf e_x}$ (black chain
curve), two valleys at $k\approx0.8k_F$ and $k\approx2k_F$ are observed in the
momentum-magnitude dependence, which correspond to the unpairing
regions mentioned above. Whereas in the pairing regions, it is found that $-\Delta^R_{\rm
  SE}({\bf k}=k{\bf e_x},\omega)$ decreases with the increase of the momentum monotonically.

The vanishing order parameter at large momentum can be understood
as follows. It has been pointed out in the previous
works\cite{Symmetry,Odd,ML-1,MLS-0,MLS-6} that the pairing function $f_{\bf k}(\omega)$ vanishes
in QW when $\xi^c_{\bf k}\gg\Delta^T_{\rm 
  SE}$. With the renormalization-induced order parameter
$\Delta({\bf k},\omega)=\int\frac{d\omega'd{\bf
    k'}}{4(2\pi)^3}\frac{V^0_{\bf k-k'}f_{\bf k'}(\omega')}{\epsilon(\omega-\omega')}$, one has
$\Delta({\bf
  k},\omega)\propto{\int}d\omega'\frac{f_{\bf k}(\omega')}{\epsilon(\omega-\omega')}$
since the Coulomb interaction $V^0_{\bf k-k'}$ is very strong at ${\bf
  k'=k}$ and can be approximated by a delta-function:
$\delta({\bf k'-k})$ in the analytic analysis. Therefore, the order parameter vanishes at large momentum 
when $\xi^c_{\bf k}\gg\Delta^T_{\rm SE}$ due to the vanishing pairing function.

This vanishing of the SE order parameter can also be understood from
Eq.~(\ref{deltas}). Specifically, after some simplifications at low
  temperature (refer to
  Appendix~\ref{fed}),  the
renormalization-induced SE order parameter can be approximately written as
\begin{eqnarray}
-\Delta^R_{\rm SE}({\bf k},\omega)&\approx&\int\frac{d{\bf
    k'}}{16\pi^2}\frac{{V^0_{\bf k-k'}}\Delta^T_{\rm
    SE}}{F_{k'}}\nonumber\\
 &&\mbox{}\times\bigg[1-\frac{\omega^{pl}_{\bf k-k'}}{F_{k'}}(1+\frac{\omega^2}{F^2_{k'}})-\frac{\xi^c_{\bf k'}\varepsilon^c_{\bf q}}{F^2_{k'}}\bigg],~~~~~~\label{SE-re}
\end{eqnarray}
where ${F_{k'}}={\sqrt{|\xi^c_{\bf k'}|^2+|\Delta^T_{\rm SE}|^2}}$.

Then, the momentum-magnitude dependence of the renormalization-induced SE order
parameter can be analyzed. Specifically, by approximately taking the Coulomb interaction
$V^0_{\bf k-k'}$ in Eq.~(\ref{SE-re}) as a delta-function $\delta({\bf k-k'})$
for the analytic analysis, one has 
$-\Delta^R_{\rm SE}({\bf k},\omega)\propto{1/\sqrt{|\xi^c_{\bf
          k}|^2+|\Delta^T_{\rm SE}|^2}}$. Then, at $\xi^c_{k}\ll\Delta^T_{\rm
  SE}\approx\xi^c_{3k_F}$ ($\xi^c_{k}\gg\Delta^T_{\rm SE}$), $-\Delta^R_{\rm
  SE}({\bf k},\omega)$ is marginally changed (decreases) with the increase of the
momentum, as shown in Fig.~\ref{figyw3}. This conclusion is  
in agreement with the one from the analysis of the pairing function proposed
above. 

Furthermore, from Eq.~(\ref{SE-re}) the CM momentum dependence of
$\Delta^R_{\rm SE}$ can also be understood. In the integral of
Eq.~(\ref{SE-re}), the third term makes the important contribution only when
$|\xi^c_{k'}|\sim{F_{k'}}$ where
$\xi^c_{k'}>\xi^c_{3k_F}$. Hence, the increase of $q$ leads to the decrease of $-\Delta^R_{\rm SE}$,
as shown in Fig.~\ref{figyw3}.

\subsection{SO ORDER PARAMETER}
\label{SO-order}
In this part, we investigate the induced SO order parameter in QW by calculating
$\Delta_s({\bf k},\omega)$ [Eq.~(\ref{deltas})] explicitly. 
Then the induced SO order parameter is obtained: 
\begin{equation}
    \label{SO}
\Delta_{\rm SO}({\bf k},\omega)=[{\Delta_{s}({\bf k},\omega)-\Delta_{s}({\bf
    k},-\omega)}]/2.
\end{equation}
For the analytic analysis, similar to the
study on the renormalization-induced SE order parameters in
Sec.~\ref{SE-order}, by taking 
some simplifications (refer to Appendix~\ref{fed}),
the SO order parameter can be approximately written as 
\begin{equation}
\Delta_{\rm SO}({\bf k},\omega)=\Delta^{K}_{\rm SO}({\bf
  k},\omega)+\Delta^{S}_{\rm SO}({\bf k},\omega), \label{SOO}
\end{equation}
with
\begin{eqnarray}
&&\Delta^{K}_{\rm SO}({\bf k},\omega)
=-\omega\int\frac{d{\bf
    k'}}{16\pi^2}{V^0_{\bf k-k'}}\Delta^T_{\rm SE}\frac{\omega^{pl}_{\bf k-k'}}{F^4_{k'}}\frac{\bf k'\cdot{q}}{m^*_c},\label{SOK}\\
&&\Delta^{S}_{\rm SO}({\bf k},\omega)
=\omega\int\frac{d{\bf
    k'}}{16\pi^2}{V^0_{\bf k-k'}}\Delta^T_{\rm SE}\frac{\omega^{pl}_{\bf
    k-k'}}{F^4_{k'}}\frac{2h_{\bf q}h_{\bf k'}\xi^c_{\bf
    k'}}{F^2_{k'}}~~~~~~\nonumber\\
&&\mbox{}~~~~~~~~~~=\omega\int\frac{d{\bf
    k'}}{16\pi^2}{V^0_{\bf k-k'}}\Delta^T_{\rm SE}\frac{\omega^{pl}_{\bf
    k-k'}}{F^4_{k'}}\Gamma_{k'}\frac{k'_xq_x}{m^*_c}.\label{SOS}
\end{eqnarray}
Here, $\Gamma_{k'}={|\gamma_D\langle{k^2_z}\rangle|^2\xi_{k'}m^*_c}/{F^2_{k'}}$.
The first term $\Delta^{K}_{\rm SO}({\bf
  k},\omega)$ in Eq.~(\ref{SOO}) comes from the broken translational symmetry by
the CM momentum of Cooper pair ($q\ne0$) in the system. Whereas the second one $\Delta^{S}_{\rm SO}({\bf
  k},\omega)$ is induced due to the coupling between the CM momentum and SOC from
the high order expansion. Hence, $\Delta^{K}_{\rm SO}({\bf
  k},\omega)$ makes the leading contribution in Eq.~(\ref{SOO}).
Additionally, it is noted that $\Delta_{\rm SO}({\bf k},\omega)$
arises from the retardation effect in the effective attractive potential $V^{\rm
  at}_{\bf k}(\omega)$, in consistent with the previous
works.\cite{SO-1,SO-2,SO-3,SO-4,SO-5,Odd}   
In the following, we show that from Eq.~(\ref{SOO}), the momentum and 
dependence of the SO order parameter can be
analyzed, in good agreement with the full numerical results.

\begin{figure}[htb]
  {\includegraphics[width=8.5cm]{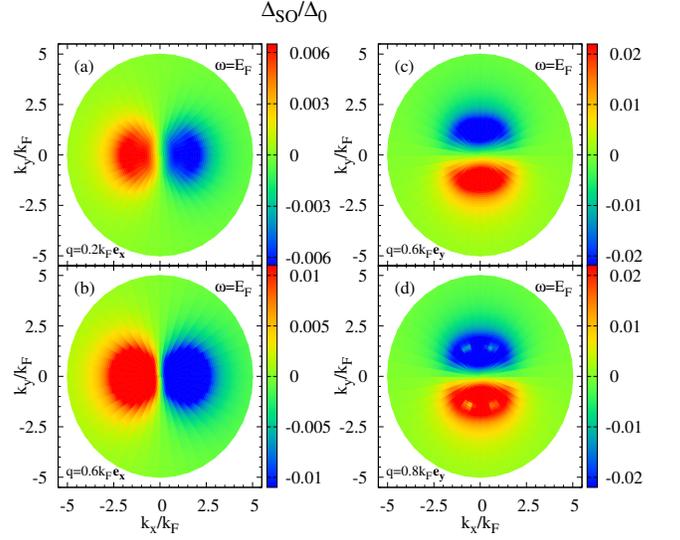}}
\caption{Momentum dependence of the SO order parameter $\Delta_{\rm SO}({\bf k},\omega=E_F)$ at (a)
${\bf q}=0.2k_F{\bf e_x}$, (b)
${\bf q}=0.6k_F{\bf e_x}$, (c)
${\bf q}=0.6k_F{\bf e_y}$ and (d)
${\bf q}=0.8k_F{\bf e_y}$. $n=5n_0$.}   
\label{figyw4}
\end{figure}

\begin{figure}[htb]
  {\includegraphics[width=9cm]{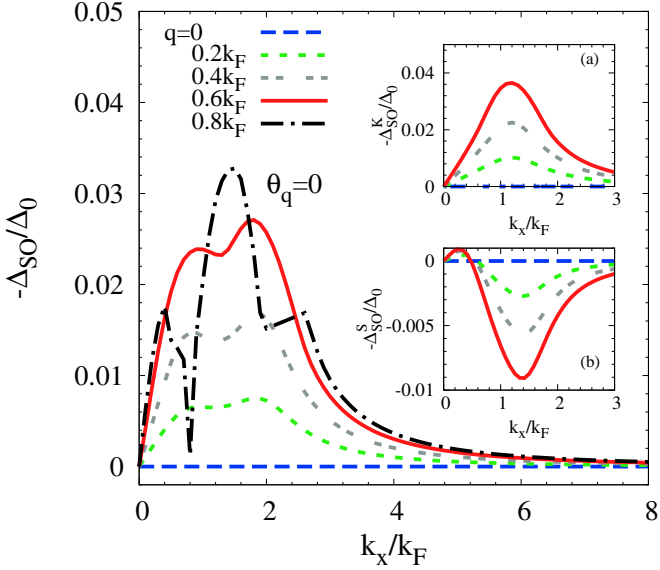}}
\caption{Momentum-magnitude dependence of the strength for the SO order
  parameter $|\Delta_{\rm SO}({\bf
    k}=k{\bf e_x},\omega=E_F)|=-\Delta_{\rm SO}({\bf
    k}=k{\bf e_x},\omega=E_F)$ at different CM momentums $q{\bf
    e_x}$.  The inset (a) [(b)] shows the calculated results for $-\Delta^K_{\rm SO}({\bf
    k}=k{\bf e_x},\omega=E_F)$ [$-\Delta^S_{\rm SO}({\bf
    k}=k{\bf e_x},\omega=E_F)$] from Eq.~(\ref{SOK}) [Eq.~(\ref{SOS})].  $n=5n_0$.}   
\label{figyw5}
\end{figure}

\subsubsection{Momentum dependence of the SO order parameter: $p$-wave character}

In this part, we focus on the momentum dependences of the SO order 
parameter $\Delta_{\rm SO}({\bf k},\omega=E_F)$ at different CM
momentums ${\bf q}$. The full numerical results are plotted in Fig.~\ref{figyw4}
at $n=5n_0$. 
As seen from the figure, it is found that the SO order parameter in the momentum
space exhibits a $p$-wave character: $\Delta_{\rm SO}=C_{\rm SO}(k,\omega){\bf k{\cdot{q}}}$ with
    $C_{\rm SO}(k,\omega)$ being independent on the orientation of the
momentum (demonstrated in Appendix~\ref{AP-md}). This 
can be understood as follows. With the Fermi-Dirac statistics for the SO order parameter, one
finds $\Delta_{\rm SO}({\bf k},\omega)=-\Delta_{\rm SO}(-{\bf
  k},\omega)$. In addition, the system in QW has the spatial-rotational
symmetry around the ${\bf q}$ axis in the momentum space, and then one has
$\Delta_{\rm SO}({\bf k},\omega)=\Delta_{\rm SO}(-{\bf k},\omega)$ when
${\bf k}\perp{\bf q}$. Hence, the vanishing $\Delta_{\rm SO}({\bf
  k},\omega)$ at ${\bf k}\perp{\bf q}$ is immediately obtained and the SO order 
parameter exhibits as $\Delta_{\rm SO}({\bf k},\omega)\propto{\bf
  k{\cdot}q}$ at small $q$. 

From Eq.~(\ref{SOO}), one can obtain the same
conclusion. Specifically, by approximately taking the Coulomb interaction
$V^0_{\bf k-k'}\omega^{pl}_{\bf k-k'}\propto{1/\sqrt{|{\bf k-k'}|}}$ as a delta-function,
one has $\Delta_{\rm SO}({\bf k},\omega)\propto{\bf k{\cdot}q}$.  Additionally, it is noted
that small strengths of the SO order parameter are observed in the unpairing
regions at $q=0.8k_F$ [shown by the four green regions
in Fig.~\ref{figyw4}(d)] due to the renormalization of the e-e Coulomb
interaction, similar to the renormalization-induced SE order parameter.

\subsubsection{Momentum-magnitude dependence of the SO order parameter}

The momentum-magnitude 
dependences of the SO order parameter are shown in
Fig.~\ref{figyw5} at different CM momentums $q{\bf e_x}$. As
seen from the figure, when $q=0$, with the translational 
symmetry, the SO order parameter $\Delta_{\rm SO}=0$ (blue dashed
curve). When $q=0.8k_F$ (black chain curve), two valleys are observed at
$0.8k_F$ and $2k_F$, exactly corresponding to the unpairing regions, similar
to the results of the renormalization-induced SE order parameter.  
When $0<q<0.8k_F$, it is shown that the strength
of the SO order parameter exhibits double-peak behavior with the increase of the
momentum. 

Specifically, 
one finds that 
$\Delta_{\rm SO}({\bf k}=0,\omega)=0$ due to the odd parity and $|\Delta_{\rm SO}({\bf 
  k},\omega)|\propto{\bf k{\cdot}q}$ increases with increasing momentum
at small $k$. As for the large momentum, due to the vanishing pairing
function at $\xi^c_{k}\gg\Delta^T_{\rm SE}$ mentioned in Sec.~\ref{SE-order}, one has $\Delta_{\rm SO}({\bf
  k},\omega)=0$. Hence, peaks of
the strength for the SO order parameter in between are expected. 
Similarly, peaks for $|\Delta^{K}_{\rm
  SO}({\bf k},\omega)|$ and $|\Delta^{S}_{\rm SO}({\bf k},\omega)|$ are also
expected, as shown in the insets (a) and (b) in Fig.~\ref{figyw5} where we
 plot $-\Delta^{K}_{\rm
  SO}({\bf k},\omega)=|\Delta^{K}_{\rm
  SO}({\bf k},\omega)|$ and $-\Delta^{S}_{\rm SO}({\bf k},\omega)=-|\Delta^{S}_{\rm
  SO}({\bf k},\omega)|$ at ${\bf q}=q{\bf
      e_x}$, respectively.   
Hence, with the peak of $-\Delta^{K}_{\rm SO}$ and
the valley of $-\Delta^{S}_{\rm
  SO}$ shown in the insets (a) and (b) of Fig.~\ref{figyw5}, respectively, the strength of the SO order
parameter $-\Delta_{\rm SO}({\bf k},\omega)=-\Delta^{K}_{\rm
  SO}({\bf k},\omega)-\Delta^{S}_{\rm SO}({\bf k},\omega)$ exhibits the
double-peak structure in the momentum-magnitude
dependence. Particularly, it is found that at the case ${\bf q}=q{\bf
      e_y}$, with $\Delta^S_{\rm SO}\approx0$, the strength of the SO order
    parameter $-\Delta_{\rm SO}({\bf k},\omega)=-\Delta^K_{\rm SO}({\bf k},\omega)$ shows
    one-peak structure (not shown in the figure).

From Eq.~(\ref{SOK}) [Eq.~(\ref{SOS})], the peak positions for $|\Delta^{K}_{\rm
  SO}({\bf k},\omega)|$ and $|\Delta^{S}_{\rm SO}({\bf k},\omega)|$ at
$q{\bf e_x}$ can be determined explicitly. Specifically, by considering the Coulomb interaction
$V^0_{\bf k-k'}\omega^{pl}_{\bf k-k'}$ as a delta-function
for the analytic analysis,  
one has $|\Delta^{K}_{\rm SO}|\propto{\bf k{\cdot}q}{/(|\xi^c_{\bf
          k}|^2+|\Delta^T_{\rm SE}|^2)^2}$ [$|\Delta^{S}_{\rm
SO}|\propto{\bf k{\cdot}q}\varepsilon^c_{k}{/(|\xi^c_{\bf
          k}|^2+|\Delta^T_{\rm SE}|^2)^3}$], exhibiting one-peak behavior in the
momentum-magnitude dependence. Then, by calculating the
maxima of ${\bf k{\cdot}q}{{{/(|\xi^c_{\bf 
          k}|^2+|\Delta^T_{\rm SE}|^2)^2}}}$ and ${\bf k{\cdot}q}\varepsilon^c_{k}{{{/(|\xi^c_{\bf
          k}|^2+|\Delta^T_{\rm SE}|^2)^3}}}$, 
the peak positions for $|\Delta^{K}_{\rm SO}|$
and $|\Delta^{S}_{\rm SO}|$ can be determined at 
${k^{p}_{K}}\approx{\sqrt{3+\sqrt{9+7(1+|\Delta^T_{\rm SE}|^2E^{-2}_F)}}}k_F/{\sqrt{7}}=1.15k_F$
and $k^{p}_{S}\approx{\sqrt{1+\sqrt{1+3(1+|\Delta^T_{\rm
          SE}|^2E^{-2}_F)}}}k_F/{\sqrt{3}}=1.26k_F$, respectively, 
very close to those from the numerical calculation shown in the
insets (a) and (b) of Fig.~\ref{figyw5}. Then, the valley position between
the two peaks in the momentum-magnitude dependence for $-\Delta_{\rm SO}({\bf
  k},\omega)$ can be determined at $k^{p}_{S}\approx1.26k_F$, close to the one from
full numerical results shown in Fig.~\ref{figyw5}.

Furthermore, due to the broken translational symmetry by the CM
momentum ${\bf q}$, the strength of the induced SO order parameter increases
with the increase of $q$ as shown in Fig.~\ref{figyw5}, in consistent with Eq.~(\ref{SOO}).

\subsection{TE ORDER PARAMETER}
\label{TE-order}
In this part, we focus on the TE order parameters in QW by numerically
calculating $\Delta_t({\bf k},\omega)$ [Eq.~(\ref{deltat})]. The induced TE order
    parameter is obtained:
\begin{equation}
    \label{TE}
\Delta_{\rm TE}({\bf k},\omega)=[{\Delta_{t}({\bf k},\omega)+\Delta_{t}({\bf
    k},-\omega)}]/2.
\end{equation}
For the analytic
analysis, similar to the
study on the SE order parameter, by taking some simplifications (refer to Appendix~\ref{fed}), the induced TE order
    parameter can be approximately written as
\begin{eqnarray}
&&\Delta_{\rm TE}({\bf k},\omega)=\int\frac{d{\bf
    k'}}{16\pi^2}\frac{{{V^0_{\bf k-k'}}\Delta^T_{\rm SE}}}{F^3_{k'}}{h_{\bf
    k'}\xi^c_{\bf k'}}\nonumber\\
&&\mbox{}\times\left[1-\frac{2\omega^{pl}_{\bf
      k-k'}}{F_{k'}}\left(1+\frac{\omega^2}{F^2_{\bf k'}}\right)-\frac{3\xi^c_{\bf k'}\varepsilon^c_{\bf q}}{F^2_{k'}}\right].~~~~~~\label{TE-re}
\end{eqnarray}
It is noted that the TE order parameter is induced due to the broken spin-rotational symmetry by the SOC ($h_{\bf k}\ne0$).
In the following, we show that from Eq.~(\ref{TE-re}), the momentum dependence of the TE order parameter can be
analyzed, in consistent with the numerical results well.

\begin{figure}[htb]
  {\includegraphics[width=8.5cm]{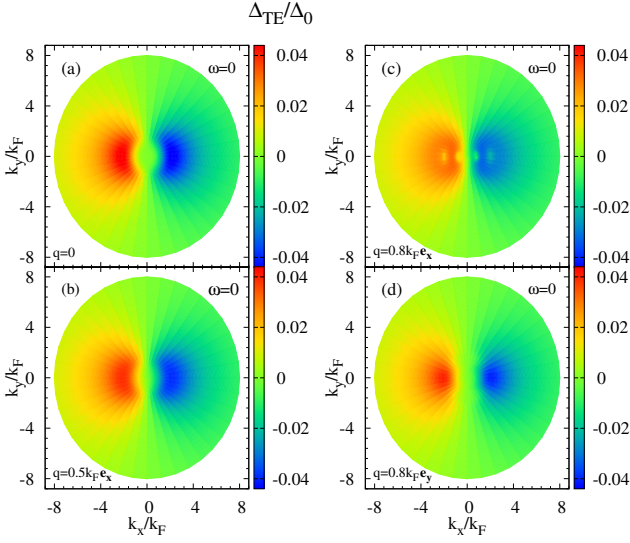}}
\caption{Momentum dependence of the
TE order parameter $\Delta_{\rm TE}({\bf k},\omega=0)$ at (a)
${\bf q}=0$, (b)
${\bf q}=0.5k_F{\bf e_x}$, (c)
${\bf q}=0.8k_F{\bf e_x}$ and (d)
${\bf q}=0.8k_F{\bf e_y}$. $n=5n_0$.}   
\label{figyw6}
\end{figure}

\begin{figure}[htb]
  {\includegraphics[width=9cm]{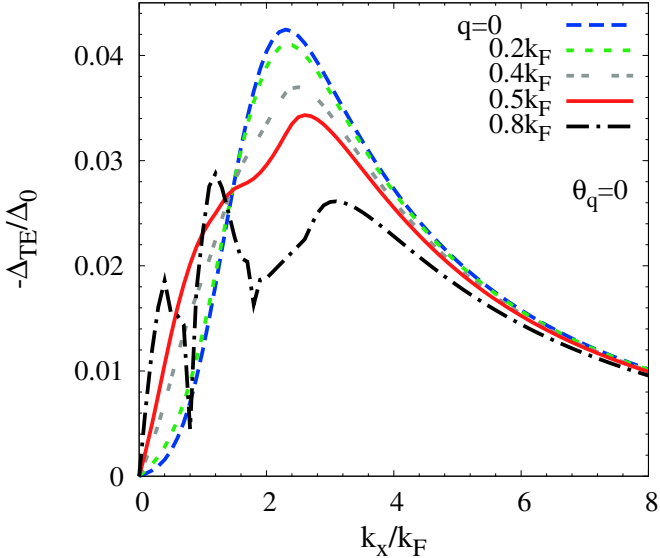}}
\caption{Momentum-magnitude dependence of the strength for the TE order
  parameter $|\Delta_{\rm TE}({\bf
    k}=k{\bf e_x},\omega=0)|=-\Delta_{\rm TE}({\bf
    k}=k{\bf e_x},\omega=0)$ at different CM momentums $q{\bf
    e_x}$.  $n=5n_0$.}   
\label{figyw7}
\end{figure}

\subsubsection{Momentum dependence of the TE order parameter: $p$-wave character} 

In this part, we show the momentum dependences of the
TE order parameter $\Delta_{\rm TE}({\bf k},\omega=0)$ at different
CM momentums ${\bf q}$. The full numerical results are 
plotted in Fig.~\ref{figyw6} at $n=5n_0$. As seen from the figure, the TE order parameter
exhibits a $p$-wave character in the momentum space: $\Delta_{\rm TE}=C_{\rm TE}(k,\omega){\bf k{\cdot{e_x}}}$ with
    $C_{\rm TE}(k,\omega)$ being independent on the orientation of the
momentum (demonstrated in Appendix~\ref{AP-md}). Differing from the $p$-wave
character of the SO order parameter which shows anisotropy with respect to the direction of ${\bf
  q}$ (as shown in Fig.~\ref{figyw4}) 
[$\Delta_{\rm SO}({\bf k},\omega)=C_{\rm SO}(k,\omega){\bf
  k{\cdot}q}$], the $p$-wave character of the TE order parameter is independent on the direction of
${\bf q}$. This is because that the TE order parameter is induced due to the
spin-rotational asymmetry by the SOC. This can
also be understood from Eq.~(\ref{TE-re}) by approximately taking the Coulomb
interaction $V^0_{\bf k-k'}$ as a delta-function, one has $\Delta_{\rm
  TE}({\bf k},\omega)\propto{h_{\bf k}}\propto{k_x}$.

Moreover, it is noted in Fig.~\ref{figyw6}(c) 
that small strengths of the TE order parameter are observed in the unpairing
regions at $q=0.8k_F$ (shown by the four green regions
with the smaller order parameters than the region nearby) due to the
renormalization of the e-e Coulomb interaction, similar to the
renormalization-induced SE and SO ones.  
  
\subsubsection{Momentum-magnitude dependence of the TE order parameter} 

The momentum-magnitude 
dependences of the TE order parameter $|\Delta_{\rm TE}({\bf
    k}=k{\bf e_x},\omega)|$ are shown in
Fig.~\ref{figyw7} at different CM momentums $q{\bf e_x}$ when $n=5n_0$. As
seen from the figure, when $q=0.8k_F$, two valleys are observed at
$0.8k_F$ and $2k_F$ in the momentum-magnitude dependence (black chain curve), 
corresponding to the unpairing regions mentioned above. 
When $q<0.8k_F$, with the increase of the momentum,
$|\Delta_{\rm TE}({\bf 
    k}=k{\bf e_x},\omega)|$ first increases when $k<2.3k_F$ then decreases after
$k>2.3k_F$, leading to a peak around $k\approx2.3k_F$.  

The peak behavior can be understood as follows. Since the TE order parameter is
induced by the SOC, with the increase of
the momentum at small (large) momentum, the TE order parameter is enhanced (suppressed) by the enhanced
SOC (suppressed pairing function mentioned in Sec.~\ref{SE-order}), leading to the peak in the
momentum-magnitude dependence. This conclusion is in good agreement with 
Eq.~(\ref{TE-re}), from which one finds that $|\Delta_{\rm TE}({\bf k},\omega)|\propto|h_{\bf k}\xi^c_{\bf
    k}|/|\sqrt{\xi_k^2+|\Delta^T_{\rm SE}|^2}|^3$ by approximately taking the Coulomb
interaction $V^0_{\bf k-k'}$ as a delta-function.
Then through the similar analysis of the momentum-magnitude peak for the
strength of the SO order
parameter in Sec.~\ref{SO-order},   
the equation of the momentum-magnitude peak position $k^p_{\rm TE}$ for the strength of the TE
order parameter can be obtained:
\begin{equation}
  \label{pTE}
3(|\frac{k^p_{\rm TE}}{k_F}|^2-1)^2(|\frac{k^p_{\rm
    TE}}{k_F}|^2+1)=2|\frac{\Delta^T_{\rm SE}}{E_F}|^2(3|\frac{k^p_{\rm
    TE}}{k_F}|^2-1). 
\end{equation}
From Eq.~(\ref{pTE}), one has $k^p_{\rm TE}\approx2.1k_F$, very close to the position from 
the full numerical calculation shown in Fig.~\ref{figyw7}. 

Moreover, the CM momentum dependence of the TE order parameter can
also be analyzed. Specifically, from Eq.~(\ref{TE-re}), by approximately taking the
Coulomb interaction $V^0_{\bf k-k'}$ as a delta-function, one finds that with
the increase of $q$, $|\Delta_{\rm TE}({\bf k},\omega)|$ increases when $k<k_F$
and decreases when $k>k_F$, in consistent with the numerical results in
Fig.~\ref{figyw7}.

\subsection{TO ORDER PARAMETER}
\label{TO-order}
In this part, we investigate the TO order parameters in QW through the numerical
calculation of $\Delta_t({\bf k},\omega)$ [Eq.~(\ref{deltat})]. The induced TO order
    parameter is obtained:
\begin{equation}
    \label{TO}
\Delta_{\rm TO}({\bf k},\omega)=[{\Delta_{t}({\bf k},\omega)-\Delta_{t}({\bf
    k},-\omega)}]/2.
\end{equation}
As for the analytic
analysis, similar to the
study on the TE order parameter in Sec.~\ref{TE-order},  by taking some
    simplifications (refer to Appendix~\ref{fed}), the TO order parameter
can approximately be written as: 
\begin{eqnarray}
\Delta_{\rm TO}({\bf k},\omega)&=&-\omega\int\frac{d{\bf
    k'}}{16\pi^2}\frac{\omega^{pl}_{\bf
    k-k'}{V^0_{\bf k-k'}}\Delta^T_{\rm SE}}{F^4_{k'}}\nonumber\\
&&\mbox{}\times\left(h_{\bf q}-\frac{\bf k'{\cdot}q}{m^*_c}\frac{h_{\bf
      k'}\xi^c_{\bf k'}}{F^2_{k'}}\right)\nonumber\\
&\approx&\omega\gamma_D\langle{k^2_z}\rangle\int\frac{d{\bf
    k'}}{16\pi^2}\frac{\omega^{pl}_{\bf
    k-k'}{V^0_{\bf k-k'}}\Delta^T_{\rm SE}}{F^4_{k'}}\nonumber\\
&&\mbox{}\times\left(q_x-\frac{\bf k'{\cdot}q}{m^*_c}\frac{k'_x\xi^c_{\bf k'}}{F^2_{k'}}\right).~~~~\label{TOE}
\end{eqnarray}
It is noted that the TO order parameter is induced due to the spin-rotational
asymmetry by the SOC ($h_{\bf k}\ne0$) and the translational asymmetry 
($q\ne0$). Additionally, the TO order parameter comes from the retardation
effect in the effective attractive potential $V^{\rm at}_{\bf k}(\omega)$, in
consistent with the previous works.\cite{TO-1,TO-2,TO-3,TO-4,TO-5,Odd}  
In the following, we show that from Eq.~(\ref{TOE}), the momentum dependence of
the TO order parameter can be 
analyzed, in good agreement with the numerical results.

\begin{figure}[htb]
  {\includegraphics[width=8.5cm]{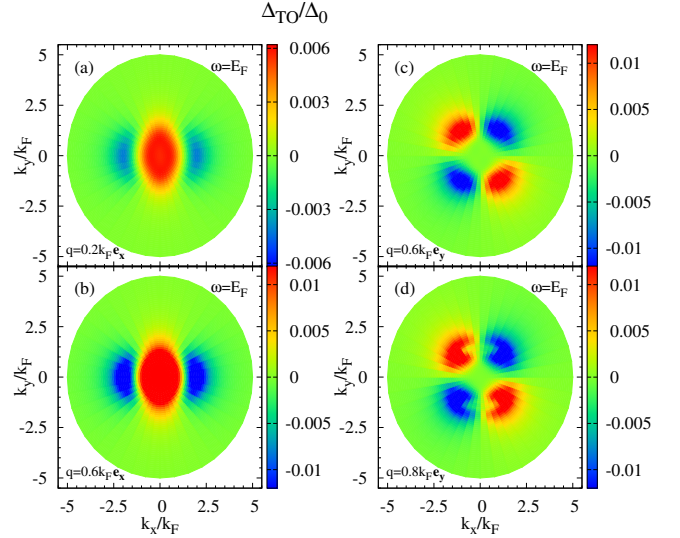}}
\caption{Momentum dependence of the
 TO order parameter $\Delta_{\rm TO}({\bf k},\omega=E_F)$ at (a)
${\bf q}=0.2k_F{\bf e_x}$, (b)
${\bf q}=0.6k_F{\bf e_x}$, (c)
${\bf q}=0.6k_F{\bf e_y}$ and (d)
${\bf q}=0.8k_F{\bf e_y}$. $n=5n_0$.}   
\label{figyw8}
\end{figure}

\begin{figure}[htb]
  {\includegraphics[width=9cm]{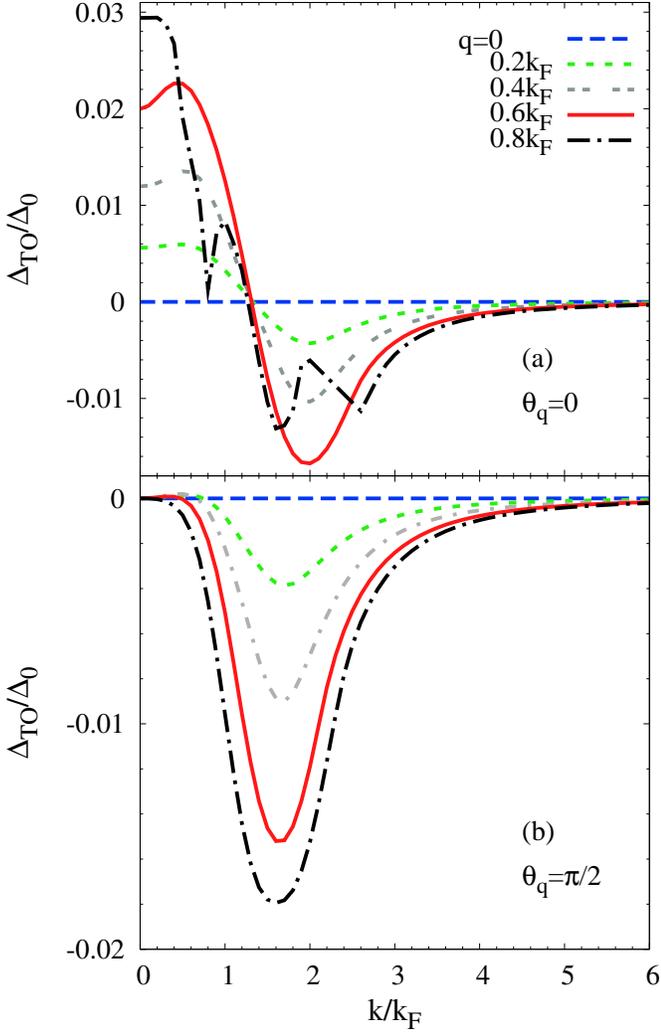}}
\caption{Momentum-magnitude dependence of the TO order
  parameter for (a) $\Delta_{\rm TO}({\bf
    k}=k{\bf e_x},\omega=E_F)$ at ${\bf 
    q}=q{\bf e_x}$ and (b) $\Delta_{\rm TO}({\bf
    k}=k{\bf e_c},\omega=E_F)$ at ${\bf q}=q{\bf e_y}$. $n=5n_0$. ${\bf e_c}$ is
the unit vector along the ${\bf e_x+e_y}$ direction.}   
\label{figyw9}
\end{figure}

\subsubsection{Momentum dependence of the TO order parameter: $d_{x^2}$ and $d_{xy}$ wave characters} 

In this part, we focus on the momentum dependences of the TO order 
parameter at different CM momentums ${\bf q}$, which are
plotted in Fig.~\ref{figyw8} at $n=5n_0$. We find that the TO order parameter shows a
$d$-wave character in the momentum space: ${\Delta_{\rm TO}({\bf
    k},\omega)={C^0_{\rm TO}(k,\omega){\bf q{\cdot}e_x}-C^1_{\rm TO}(k,\omega)({\bf
      k{\cdot}e_x})({\bf k{\cdot}q})}}$ with $C^0_{\rm TO}(k,\omega)$ and
$C^1_{\rm TO}(k,\omega)$ being independent on the orientation of the
momentum (demonstrated in Appendix~\ref{AP-md}). Particularly, at
${\bf q}=q{\bf e_x}$ (${\bf q}=q{\bf e_y}$), the TO order parameters exhibit a
$d_{x^2}$-wave ($d_{xy}$-wave) character: $\Delta_{\rm TO}({\bf k},\omega)=(C^0_{\rm TO}-C^1_{\rm
      TO}k_x^2)q$ [$\Delta_{\rm TO}({\bf k},\omega)=-C^1_{\rm
      TO}k_xk_yq$], as shown in Fig.~\ref{figyw8}(b) [(c)]. This is due to the
unique SOC in InSb (110) QW. Specifically, by approximately taking the Coulomb interaction $V^0_{\bf
  k-k'}\omega^{pl}_{\bf k-k'}$ as a
delta-function in Eq.~(\ref{TOE}), one has $\Delta_{\rm TO}({\bf
    k},\omega)\propto[q_x-k_x({\bf k{\cdot}q})\xi^c_{k}/(m^*_cF^2_k)]$, in good
  agreement with the full numerical results.

Additionally, it is noted
that small strengths of the TO order parameter are observed in the unpairing
regions at $q=0.8k_F$ [four concave parts shown in Fig.~\ref{figyw8}(d)] due to
the renormalization of the e-e Coulomb interaction, similar to the
renormalization-induced SE, SO and TE ones.   

\subsubsection{Momentum-magnitude dependence of the TO order parameter} 

The momentum-magnitude 
dependences of the TO order parameter are shown in
Fig.~\ref{figyw9} at different CM momentums ${\bf q}$ when $n=5n_0$.
We first discuss the case at ${\bf q}=q{\bf e_x}$.  It is noted that 
at $q=0.8k_F$, a valley and a peak of the TO order parameter in the
momentum-magnitude dependence are observed at $0.8k_F$ and $2k_F$ [black chain
curve in Fig.~\ref{figyw9}(a)], respectively, which exactly
correspond to the unpairing regions mentioned above, similar to the previous
results of SE, SO and TE ones. In addition, as shown in
Fig.~\ref{figyw9}(a) at ${\bf q}=q{\bf e_x}$, 
when $q<0.8k_F$, with the increase 
of the momentum, the TO order parameter $\Delta_{\rm TO}({\bf
    k}=k{\bf e_x},\omega)$, possessing same sign with $\Delta_0$ at
$k=0$,  first increases when
$k<0.5k_F$ then decreases when $0.5<k_F<2k_F$, leading to a peak
around $0.5k_F$. Moreover, when $0.5<k_F<2k_F$, it is noted that $\Delta_{\rm TO}({\bf
    k}=k{\bf e_x},\omega)$ has a sign change at
$1.4k_F$, and shows opposite sign against $\Delta_0$ when
$k>1.4k_F$. With further increasing the momentum when $k>2k_F$,
the TO order parameter tends to zero, leading to a valley around $2k_F$.   

This momentum-magnitude dependence of the TO order parameter is more
complex than the previous SE, SO and TE ones. Nevertheless, we show that
from Eq.~(\ref{TOE}), this dependence can be well understood. Specifically, by
approximately taking the Coulomb interaction $V^0_{\bf 
  k-k'}\omega^{pl}_{\bf k-k'}$ as a
delta-function, at ${\bf q}=q{\bf e_x}$, one has
\begin{equation}
 \label{TOtrend} 
\Delta_{\rm TO}({\bf
    k}=k{\bf
  e_x},\omega=E_F)\propto[1+\varepsilon^c_k(E_F-\varepsilon^c_k)/F^2_k]/F^4_k,
\end{equation}
with same sign to $\Delta_0$ at $k=0$. When ${k\ll3k_F}$ with
$F_k\approx\Delta^T_{\rm SE}$, from Eq.~(\ref{TOtrend}), it is found that
the increase of the momentum at $k<0.5k_F$ ($k>0.5k_F$)
leads to the increase
(decrease) of $\Delta_{\rm TO}({\bf 
    k}=k{\bf
  e_x},\omega)$ and the peak around $0.5k_F$. Moreover, when $0.5k_F<k\ll3k_F$, it is noted that the
TO order parameter $\Delta_{\rm TO}({\bf k}=k{\bf e_x},\omega)$ has a sign
change at $k=(1+|\Delta^T_{\rm
  SE}/E_F|^2)^{1/4}k_F\approx1.42k_F$ where one has $\Delta_{\rm TO}({\bf
    k}=k{\bf
  e_x},\omega=E_F)=0$ in Eq.~(\ref{TOtrend}), and then $\Delta_{\rm TO}({\bf
  k}=k{\bf e_x},\omega)$ is in opposite sign against $\Delta_0$
at $k>1.42k_F$. With further increasing the momentum, the suppression of
the pairing function, i.e., the increase of $F_k$, leads to the suppression on
the TO order parameter, and hence $\Delta_{\rm TO}({\bf
  k}=k{\bf e_x},\omega)$ tends to zero, leading to the valley
observed. The valley position can be determined by calculating the minimum of 
$\Delta_{\rm TO}({\bf
  k}=k{\bf e_x},\omega)$ at $k^v_{\rm TO}\approx(3+|\Delta^T_{\rm
  SE}/E_F|^2+|\Delta^T_{\rm SE}/E_F|\sqrt{3+|\Delta^T_{\rm
    SE}/E_F|^2})k_F/3=1.9k_F$, again very close to the one from the full
numerical results in Fig.~\ref{figyw9}(a).

For the case at ${\bf q}=q{\bf e_y}$, it is found that $\Delta_{\rm TO}({\bf
    k}=k{\bf e_c},\omega)\propto\varepsilon^c_k(E_F-\varepsilon^c_k)/F^6_k$ (${\bf e_c}$ is defined as the unit vector along ${\bf e_x+e_y}$ direction),
similar to Eq.~(\ref{TOtrend}) at ${\bf q}=q{\bf e_x}$. Consequently, by using the
similar analysis, one may understand the behavior in Fig.~\ref{figyw9}(b) well.

Furthermore, with the translational asymmetry broken by the CM
momentum ${\bf q}$, it is found that the strengths of the induced TO order
parameters $|\Delta_{\rm TO}({\bf k},\omega)|$ increase with the increase of the CM
momentum from Eq.~(\ref{TOE}), as shown in Fig.~\ref{figyw9}.

\begin{figure}[htb]
  {\includegraphics[width=9cm]{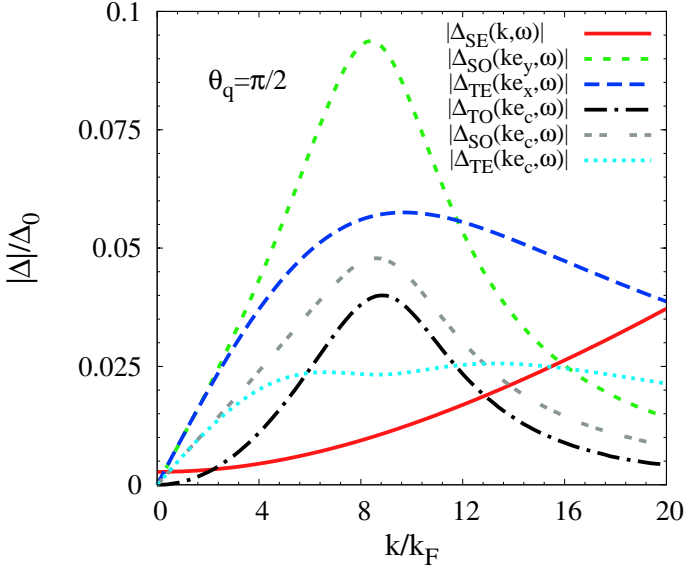}}
\caption{Momentum-magnitude dependence of the order
  parameter of each symmetry containing all four types at
  ${\bf q}=10k_F{\bf e_y}$.  It is also found that $\Delta_{\rm SO}({\bf k}=k{\bf
        e_x},\omega)=\Delta_{\rm TE}({\bf k}=k{\bf e_y},\omega)=\Delta_{\rm
        TO}({\bf k}=k{\bf e_x},\omega)=\Delta_{\rm TO}({\bf 
        k}=k{\bf e_y},\omega)=0$ from the full numerical results at
  ${\bf q}=10k_F{\bf e_y}$ (not shown in the figure). $n=0.15n_0$ and $\omega=15E_F\ll\Delta^T_{\rm
    SE}$.} 
\label{figyw10}
\end{figure}

\subsection{SEPARATION OF THE ORDER PARAMETERS CONTAINING ALL FOUR TYPES} 
\label{lalala}
Finally, we compare the four types of the order parameters 
in QW and propose a tentative way to distinguish these order parameters
in the experiment.
Specifically, through
the study of the density dependences of the four types of the order
parameters in detail (refer to Appendix~\ref{OPDD}), it is found that at small density, due to the
suppressed attractive potential $V^{\rm at}_{\bf k}$, the
renormalization-induced SE order parameter $\Delta^R_{\rm SE}$, with its sign
being opposite against the
proximity-induced one $\Delta^T_{\rm SE}$, is markedly enhanced. Then the SE
order parameter $\Delta_{\rm SE}=\Delta^T_{\rm SE}+\Delta^R_{\rm SE}$ is
efficiently suppressed at small density, similar to the previous
    work.\cite{MLS-6} In this case, the SO (induced by CM momentum of Cooper
pair), TE (induced by the SOC) and TO (induced by the SOC and CM momentum) order parameters
are dominant.  

Furthermore, as summarized in Table~\ref{Torder}, the SO
(TE) order parameter in the momentum space exhibits a $p$-wave
character while the TO one exhibits a $d$-wave
character. Particularly, when ${\bf q}=q{\bf e_y}$, from Table~\ref{Torder},
    one has $\Delta_{\rm SO}({\bf k},\omega)=C_{\rm 
  SO}(k,\omega)qk_y$, $\Delta_{\rm TE}({\bf k},\omega)=C_{\rm
  TE}(k,\omega)k_x$ and ${\Delta_{\rm TO}({\bf
    k},\omega)=-C^1_{\rm TO}(k,\omega)k_xk_y}$. In this case, with the suppressed SE order
parameter at small density, along the ${\bf e_y}$ (${\bf
  e_x}$) direction in the momentum space with $\Delta_{\rm TE}=\Delta_{\rm
  TO}=0$ ($\Delta_{\rm SO}=\Delta_{\rm
  TO}=0$), the SO (TE) order parameter can be
detected solely and then $C_{\rm SO}(k,\omega)$ [$C_{\rm TE}(k,\omega)$] can be
determined explicitly. 

The full numerical results of the
four types of the order parameters at small density $n=0.15n_0$ are plotted
against the momentum $k$ in
Fig.~\ref{figyw10} when ${\bf q}=10k_F{\bf e_y}$. In this case, due to
the small density and hence small $q$, the unpairing regions mentioned in
Sec.~\ref{SE-order} is absent here. Moreover, as seen from the figure, the SE order
parameter $\Delta_{\rm SE}$ (red solid curve) is
efficiently suppressed, and as expected above, along the ${\bf e_x}$
    (${\bf e_y}$) direction with $\Delta_{\rm SO}=\Delta_{\rm
  TO}=0$ ($\Delta_{\rm TE}=\Delta_{\rm
  TO}=0$)
the TE (SO) order parameter (blue long-dashed curve) [(green short-dashed
curve)] is dominant, providing the possibility to
distinguish this order parameter solely by detecting
along special direction.

Nevertheless,  as for the $d_{xy}$-wave TO order parameter at ${\bf
  q}=q{\bf e_y}$, along ${\bf e_c}$
direction, the maximum of the strength of the TO order parameter (black chain curve) is
comparable to that of the SO (gray double-dotted curve) and TE (light-blue
dotted curve) ones. However, with the experimentally obtained order parameter
$\Delta^e({\bf k},\omega)$, since the SE order parameter is suppressed at small 
density and the SO and TE ones can be obtained above, one obtains $\Delta_{\rm
  TO}({\bf k},\omega)=\Delta^e({\bf k},\omega)-\Delta_{\rm SO}({\bf
  k},\omega)-\Delta_{\rm TE}({\bf k},\omega)$ or
\begin{equation}
  \Delta_{\rm
  TO}({\bf k},\omega)=\int{d\theta_{\bf k}}\Delta^e({\bf
  k},\omega){\rm sin}\theta_{\bf k}{\rm cos}\theta_{\bf k}/(4\pi).
\end{equation}
Then, the $d_{xy}$-wave TO order parameter can also
    be detected. Moreover, with the strength comparable to the SO one, the
TO order parameter has been shown to provide significant protection to the zero-energy states
including the Andreev bound state and Majorana fermion state due to the even parity,\cite{Symmetry} and
promises to lead to rich physics\cite{ML-1,a-6,a-7} including the long-range proximity
effect\cite{ML-1} and anomalous features of quasiparticle.\cite{a-6,a-7}

\section{SUMMARY AND DISCUSSION}
\label{summary}
In summary, we have demonstrated that the SE, SO, TE and TO pairings and the
corresponding order
parameters can be realized in spin-orbit-coupled InSb (110) QW in proximity
to $s$-wave superconductor in FFLO phase or with a supercurrent. Specifically, the SE order parameter can 
be induced in QW through the proximity effect. Nevertheless, it is found that
there exist unpairing regions in the momentum space, in which the
proximity-induced SE order parameter vanishes. We further reveal that the
unpairing regions arise from the FFLO-phase-like blocking in QW. In the presence
of this proximity-induced SE order parameter, the SO pairing is induced due to
the broken translational symmetry by the CM momentum of Cooper pair whereas the TE one is induced
due to the broken spin-rotational symmetry by the SOC. Moreover, with the translational
and spin-rotational asymmetries, the TO pairing is induced. Then, the
corresponding order parameters can be induced from the self-energy of the e-e Coulomb
interaction with the dynamic screening, and the proximity-induced SE order
parameter is also renormalized. Particularly, we reveal that the odd-frequency order 
parameters are induced due to the retardation effect of the Coulomb interaction
from the dynamic screening in 2DEG, where the plasmon effect is important.
  
Differing from the vanishing proximity-induced SE order parameter in the unpairing 
regions, we show that through the renormalization, the SE, SO, TE and TO
order parameters in QW all have small strengths in the unpairing
regions. In the pairing regions, the SE and TE order parameters are revealed to 
show the conventional $s$-wave and $p$-wave characters, respectively. As for
the odd-frequency order parameters, which are difficult to realize in
bulk superconductors, the induced SO and TO order parameters in InSb (110) QW
exhibit the $p$-wave and $d$-wave characters, respectively. Specifically, with the broken
translational symmetry by the CM momentum of Cooper pair, 
the $p$-wave character of the SO order parameter shows anisotropy with
respect to the direction of the CM momentum. This is very different from the TE one,
which is determined by the SOC and hence is independent on the direction of the
CM momentum. As for the unconventional $d$-wave TO order parameter, it is
interesting to find that $d_{x^2}$-wave and $d_{xy}$-wave TO order parameters
can be obtained when the CM momentum is along the $[1{\bar 1}0]$ and $[001]$
directions, respectively. It is further demonstrated that this anisotropy of the
TO order parameter arises from the unique SOC structure in InSb (110) QW.

Furthermore, we show
that at proper density, the SE order parameter can be
efficiently suppressed. Then, the induced SO, TE and TO order parameters
can be detected experimentally. Our work
provides an
idea platform where rich physics, including the enhanced Josephson
current,\cite{a-1} dispersionless zero-energy Andreev bound
states\cite{Symmetry,He-2,ad-2,ad-3,ad-4} and anomalous proximity 
effect\cite{Symmetry,a-20,a-30} related to the SO order parameter; the conventional 
triplet superconductivity\cite{TE-1,TE-2,TE-3,a-40,a-50,a-20,a-30} associated
with the TE one; the long-range proximity effect\cite{ML-1} and anomalous features
of quasiparticle\cite{a-6,a-7} due to the TO one, can be realized.

\begin{acknowledgments}
This work was supported by the National Natural Science Foundation of
China under Grant No.\ 11334014 and \ 61411136001, and the Strategic Priority
Research Program of the Chinese Academy of Sciences under Grant No.\
XDB01000000.

\end{acknowledgments}

\begin{appendix}
\begin{figure}[htb]
  {\includegraphics[width=7cm]{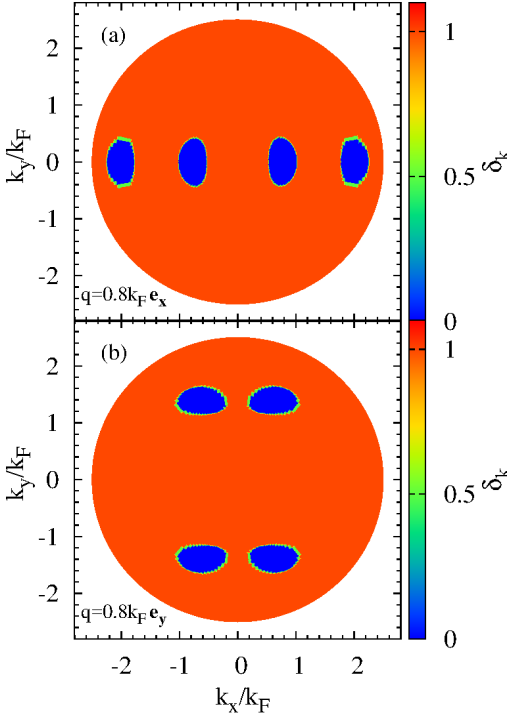}}
\caption{Momentum dependence of the depairing operator at (a)
  ${\bf q}=0.8k_F{\bf e_x}$ and (b) ${\bf q}=0.8k_F{\bf e_y}$. $n=5n_0$.}
\label{figyw11}
\end{figure}

\begin{figure}[htb]
  {\includegraphics[width=8cm]{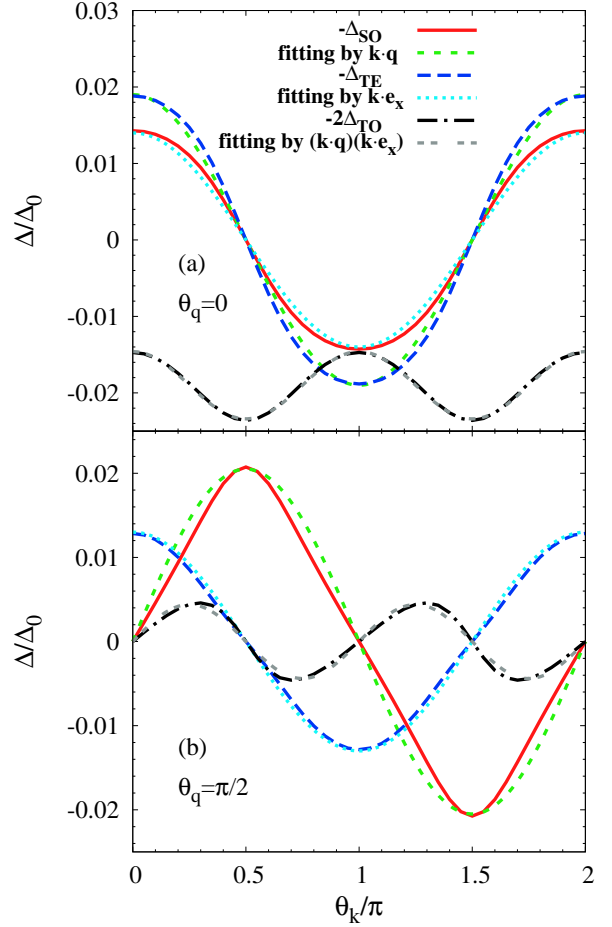}}
\caption{Momentum-orientation dependence of the order parameters at (a)
  ${\bf q}=0.4k_F{\bf e_x}$ and (b) ${\bf q}=0.4k_F{\bf e_y}$. $n=5n_0$. $\omega=E_F$ and
  $k=k_F$. Red solid, blue long-dashed and black chain curves denote the full
  numerical results of the SO, TE and TO order parameters, respectively.
  Green short-dashed curve (Light-blue dotted curve): fitted
  results for the SO (TE) order parameter by using $\Delta_{\rm SO}=C_{\rm
  SO}(k,\omega){\bf k{\cdot}q}$ [$\Delta_{\rm
  TE}=C_{\rm TE}(k,\omega){\bf k{\cdot}e_x}$]. Grey double-dotted curve:
  fitted results for the TO one by using $\Delta_{\rm TO}=C^0_{\rm
  TO}(k,\omega)-C^1_{\rm TO}(k,\omega)({\bf k{\cdot}e_x})({\bf k{\cdot}q})$.
}
\label{figyw12}
\end{figure}

\section{DERIVATION OF EQ.~(\ref{ggll})}
\label{AP-de}

We derive Eq.~(\ref{ggll}) in this part. By neglecting the frequency
dependence of the self-energy due to the proximity effect as mentioned in Sec.~\ref{sdpe}, 
Eq.~(\ref{amp}) can be expressed as
\begin{equation}
\label{Apd1}
\frac{{\hat \Delta_T}({\bf
  k})}{\Delta_0}=\int_{|\varepsilon^s_{\bf
    k{\pm}q}+\xi^s_{p_z}|<\omega_D}\frac{m^*_sd\varepsilon^s_{p_z}}{2{\pi}p_z}\left(\begin{array}{cc}  
0 & \frac{|t_0|^2}{{\Xi}^+_{{\bf k},p_z}}\\
\frac{|t_0|^2}{-{\Xi}^-_{{\bf k},p_z}}& 0\\
\end{array}\right),
\end{equation}
where $\omega_D$ represents the Debye
frequency of superconductor and ${\Xi}^{\pm}_{\bf k}(\xi^s_{p_z})=(\varepsilon^s_{\bf
  k+q}+\xi^s_{p_z}\pm{h_B})(\varepsilon^s_{\bf
  k-q}+\xi^s_{p_z}\mp{h_B})+|\Delta_0|^2$. 

Moreover, with the small momentum ${\bf 
  k}$ in QW, one has ${\varepsilon^s_{\bf k}}\ll\omega_D\ll{\mu_s}$ and hence
the restriction ${|\varepsilon^s_{\bf
    k{\pm}q}+\xi^s_{p_z}|<\omega_D}$ in the integral of Eq.~(\ref{Apd1}) can
be approximated as ${|\xi^s_{p_z}|<\omega_D}$. Then by using
the mean value theorem for the integral in Eq.~(\ref{Apd1}), one gets
\begin{equation}
{\hat \Delta_T}({\bf
  k})=\frac{|t_0|^2\Delta_0\omega_D\sqrt{m^*_s}}{\pi\sqrt{2\mu_s}}\left(\begin{array}{cc}  
0 & \frac{1}{L^+_{\bf k}}\\
-\frac{1}{L^-_{\bf k}}& 0\\
\end{array}\right),
\end{equation}
with $L^{\pm}_{\bf k}=\varepsilon^s_{\bf k+q}\varepsilon^s_{\bf
  k-q}-h_B^2+|\Delta_0|^2\pm(\varepsilon^s_{\bf k+q}-\varepsilon^s_{\bf
  k-q})h_B$. 

As mentioned in Sec.~\ref{SE-order}, it is found that
$|\Delta_0|\approx\varepsilon^s_{15k_F}$, and hence $(\varepsilon^s_{\bf
  k+q}-\varepsilon^s_{\bf k-q})$ in $L^{\pm}_{\bf k}$ can be neglected due to the
small CM momentum $q$ in this work. Then, Eq.~(\ref{ggll}) is obtained with the
effective tunneling matrix element $|{\tilde t}|^2=|t_0|^2\omega_D\sqrt{m^*_s/(2\mu_s)}/\pi$.

\section{MOMENTUM DEPENDENCE OF THE DEPAIRING OPERATOR}
\label{AP-mo}

In this part, we show the momentum dependences of the depairing operator
$\delta_{\bf k}$ at different CM momentums, which are plotted in
Fig.~\ref{figyw11}. We find that at $n=5n_0$, when $q>0.74k_F$, there always
exist zero-value regions of
the depairing operator in the momentum space (shown by the blue regions in
Fig.~\ref{figyw11} at $q=0.8k_F$), which exactly correspond to the 
unpairing regions in Sec.~\ref{SE-order} [blue regions in Figs.~\ref{figyw1}(c)
  and~(d)].

Nevertheless, due to the existence of the SOC, there exist four unpairing regions
in the momentum space at fixed ${\bf q}$ when $q>0.74k_F$. This is very
different from the conventional FFLO superconductor,\cite{FF,LO} where
only two unpairing regions exist. This can also be understood from
the four non-degenerate quasi-particle energy spectra addressed in
Sec.~\ref{SPEC}, differing from the two double-degenerate quasi-particle energy
spectra in the conventional FFLO superconductors.\cite{FF,LO}

\section{DERIVATION OF EQS.~(\ref{SE-re}), (\ref{SOO}), (\ref{TE-re}) and~(\ref{TOE})}
\label{fed}
We derive Eqs.~(\ref{SE-re}), (\ref{SOO}), (\ref{TE-re}) and~(\ref{TOE}) in this
part. Specifically, at low 
temperature, one has $f(E^{\mu}_{\nu{\bf k}})\approx0$ since $E^{\mu}_{\nu{\bf k}}>0$ and $n(\omega^{pl}_{\bf 
  k})\omega^{pl}_{\bf k}\approx0$, so the second and the third terms in
Eq.~(\ref{deltas}) can be neglected. Moreover, since the Coulomb 
interaction $V^0_{\bf k-k'}$ is very strong at ${\bf k'=k}$, the plasma
frequency $\omega^{pl}_{\bf k-k'}\propto\sqrt{|{\bf k-k'}|}$ in the denominator
of the first term can also be neglected (compared with $E^{\mu}_{\nu{\bf
    k'}}$). Then, the renormalization-induced singlet order parameter
[Eq.~(\ref{deltas})] is simplified into   
\begin{equation}
\Delta_{\rm s}({\bf k},\omega)=\sum_{\mu=\pm}\int\frac{d{\bf
      k'}}{32\pi^2}{{V^0_{\bf k-k'}}\Delta^T_{\rm SE}}\bigg(\frac{\omega^{pl}_{\bf k-k'}}{A^{\mu+}_{\bf
      k'}}-\frac{1}{B^{\mu}_{+{\bf
          k'}}}\bigg),
\end{equation}
where $A^{\mu\nu}_{\bf k'}=E^{\mu}_{+{\bf k'}}E^{\mu}_{-{\bf
    k'}}-\omega^2+\nu{\omega}B^{\mu}_{-{\bf
        k'}}$ and $B^{\mu}_{\pm{\bf k'}}=E^{\mu}_{+{\bf k'}}{\pm}E^{\mu}_{-{\bf
    k'}}$. 

Then, the renormalization-induced SE order parameter and SO one can be obtained
from Eqs.~(\ref{SE}) and~(\ref{SO}), respectively, written as 
\begin{equation}
\Delta^R_{\rm SE}({\bf k},\omega)=\sum_{\nu,\mu=\pm}\int\frac{d{\bf
      k'}}{64\pi^2}{{V^0_{\bf k-k'}}\Delta^T_{\rm SE}}\bigg(\frac{\omega^{pl}_{\bf k-k'}}{A^{\mu\nu}_{\bf
      k'}}-\frac{1}{B^{\mu}_{+{\bf
          k'}}}\bigg),\label{deltas-sims}~~~~~~ 
\end{equation}
\begin{equation}
\label{SOO-sims}
\Delta_{\rm SO}({\bf k},\omega)=\sum_{\nu,\mu=\pm}\nu\int\frac{d{\bf
      k'}}{64\pi^2}\frac{{{V^0_{\bf k-k'}}\Delta^T_{\rm SE}}\omega^{pl}_{\bf k-k'}}{A^{\mu\nu}_{\bf
      k'}}. 
\end{equation}
By approximately considering $\omega$,
$\varepsilon^c_{\bf q}$ and $h_{\bf k'}$ as small quantities (compared with
$\xi^c_{\bf k'}$ and ${\Delta}^T_{\rm SE}$) for expansion,   
Eq.~(\ref{deltas-sims}) [Eq.~(\ref{SOO-sims})] can be further simplified, and
then Eq.~(\ref{SE-re}) [Eq.~(\ref{SOO})] can be obtained.

Similarly, by taking the same approximation $n(\omega^{pl}_{\bf
    k})\omega^{pl}_{\bf k}\approx0$ and
  $f(E^{\mu}_{\nu{\bf k}})\approx0$ at low temperature
and $\omega^{pl}_{\bf  
  k-k'}<E^{\mu}_{\nu{\bf k'}}$ in Eq.~(\ref{deltat}), the
induced TE order parameter and TO one can be obtained 
from Eqs.~(\ref{TE}) and~(\ref{TO}), respectively, written as 
\begin{equation}
\Delta_{\rm TE}({\bf k},\omega)=\sum_{\nu,\mu=\pm}\mu\int\frac{d{\bf
      k'}}{64\pi^2}{{V^0_{\bf k-k'}}\Delta^T_{\rm SE}}\bigg(\frac{\omega^{pl}_{\bf k-k'}}{A^{\mu\nu}_{\bf
      k'}}-\frac{1}{B^{\mu}_{+{\bf
          k'}}}\bigg),\label{deltat-sims}
\end{equation}
\begin{equation}
\Delta_{\rm TO}({\bf k},\omega)\sum_{\nu,\mu=\pm}\mu\nu\int\frac{d{\bf
      k'}}{64\pi^2}\frac{{{V^0_{\bf k-k'}}\Delta^T_{\rm SE}}\omega^{pl}_{\bf k-k'}}{A^{\mu\nu}_{\bf
      k'}}.\label{deltat-simss}~~~~~~ 
\end{equation}
Then, by further approximately taking $\omega$,
$\varepsilon^c_{\bf q}$ and $h_{\bf k'}$ as small quantities, 
Eq.~(\ref{deltat-sims}) [Eq.~(\ref{deltat-sims})] is simplified into
Eq.~(\ref{TE-re}) [Eq.~(\ref{TOE})].

\section{MOMENTUM-ORIENTATION DEPENDENCE OF ORDER PARAMETERS}
\label{AP-md}

In this part, we show the momentum-orientation dependences of the four order
parameters at different CM momentums, which 
are plotted in Fig.~\ref{figyw12} at $n=5n_0$. As seen from the figure, 
the full
numerical results of the SO (red solid curve), TE (light-blue dotted
curve) and TO (gray double-dotted curve) order parameters can be well fitted
by using
the analytic analyses addressed in Table~\ref{Torder}: $\Delta_{\rm SO}=C_{\rm
  SO}(k,\omega){\bf k{\cdot}q}$ (green short-dashed curve), $\Delta_{\rm
  TE}=C_{\rm 
  TE}(k,\omega){\bf k{\cdot}e_x}$ (blue long-dashed curve) and  $\Delta_{\rm TO}=C^0_{\rm
  TO}(k,\omega)-C^1_{\rm TO}(k,\omega)({\bf k{\cdot}e_x})({\bf k{\cdot}q})$
(black chain curve), respectively. 

\begin{figure}[htb]
  {\includegraphics[width=9cm]{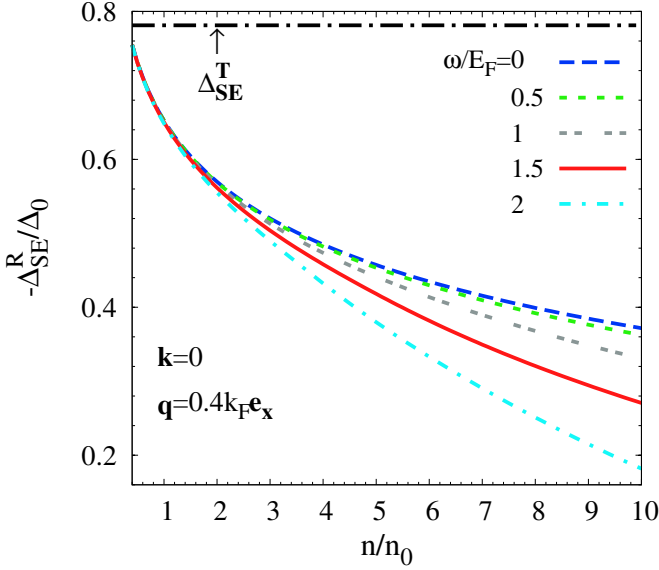}}
\caption{Strength of the renormalization-induced SE order parameter $|\Delta^R_{\rm SE}({\bf
  k=0},\omega)|=-\Delta^R_{\rm SE}({\bf
  k=0},\omega)$
  versus density $n$ at different frequencies $\omega$. Black chain line:
  results for
  $\Delta^T_{\rm SE}$. ${\bf q}=0.4k_F{\bf e_x}$}    
\label{figyw13}
\end{figure}

\section{DENSITY DEPENDENCES OF THE ORDER
PARAMETERS CONTAINING ALL FOUR
TYPES}
\label{OPDD}
\subsection{Density dependence of the renormalization-induced SE order parameter} 
 
We first address the density dependence of the  
renormalization-induced SE order parameter.  The strengths of the 
renormalization-induced SE order parameter $-\Delta^R_{\rm
  SE}({\bf k}=0,\omega)$ versus density $n$ when ${\bf q}=0.4k_F{\bf e_x}$ are plotted
in Fig.~\ref{figyw13} at different frequencies $\omega$.  
As shown from the figure, $-\Delta^R_{\rm
  SE}({\bf k}=0,\omega)$ decreases monotonically with the increase of the
density. This behavior can also be understood from Eq.~(\ref{SE-re}). The vanishing
renormalization-induced SE order parameter at large density comes from the
vanishing pairing function at $|\xi^c_{k=0}|\gg\Delta^T_{\rm SE}$ as mention
above, similar to the case at large momentum. At small density, the
attractive potential $V^{\rm at}_{\bf k}$ is efficiently suppressed, leading to
the enhanced Coulomb interaction and hence the renormalization-induced
SE order parameter.

Furthermore, it is noted that with the decrease of the density, the strength of the 
renormalization-induced SE order parameter $-\Delta^R_{\rm
  SE}$ becomes close to that of the proximity-induced one
$\Delta^T_{\rm SE}$ (black chain line). Since  $\Delta^R_{\rm
  SE}$ is always in the opposite sign against $\Delta^T_{\rm SE}$, the SE
order parameter $\Delta_{\rm SE}=\Delta^T_{\rm SE}+\Delta^R_{\rm SE}$ in QW can be
efficiently suppressed, 
providing the possibility for the experimental observation for SO, TE and TO
ones, as mentioned in Sec.~\ref{lalala}.

Additionally, from Fig.~\ref{figyw13}, it is found that at fixed density, the
strength of the renormalization-induced SE order parameter  $-\Delta^R_{\rm
  SE}({\bf k}=0,\omega)$ decreases monotonically with the increase of the frequency,
in consistent with the analytic results in Eq.~(\ref{SE-re}).

\begin{figure}[htb]
  {\includegraphics[width=9cm]{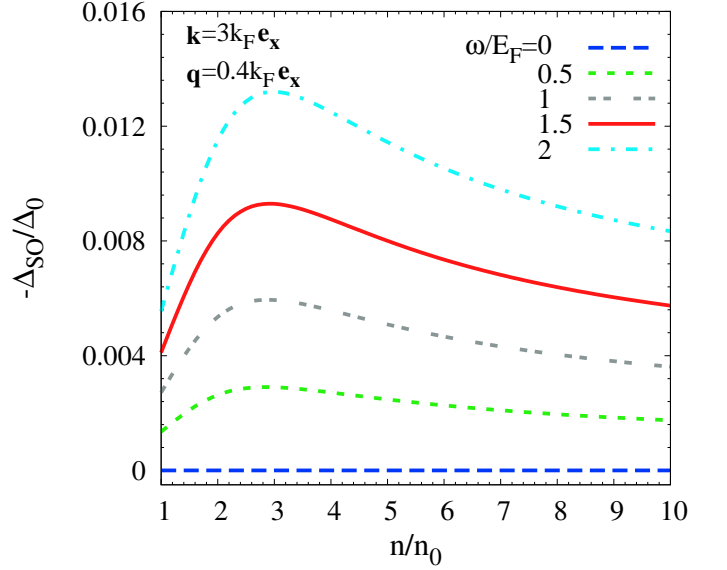}}
\caption{Strength of the SO order parameter $|\Delta_{\rm
  SO}({\bf k}=3k_F{\bf e_x},\omega)|=-\Delta_{\rm
  SO}({\bf k}=3k_F{\bf e_x},\omega)$ versus density $n$ at different
frequencies. ${\bf q}=0.4k_F{\bf e_x}$.}   
\label{figyw14}
\end{figure}

\subsection{Density dependence of the SO order
  parameter} 

We next address the density dependence of the
SO order parameter.  The strengths of the
SO order parameter $|\Delta_{\rm
  SO}({\bf k}=3k_F{\bf e_x},\omega)|$ as function of the density $n$ are
plotted in Fig.~\ref{figyw14} at different
frequencies when ${\bf q}=0.4k_F{\bf e_x}$. 
As seen from the figure, with the increase of the density $n$, $|\Delta_{\rm
  SO}({\bf k}=3k_F{\bf e_x},\omega)|$ first increases when $n<2.8n_0$ then decreases when
$n>2.8n_0$, leading to a peak around $2.8n_0$.

The peak behavior can be understood as follows. At small (large) density with $\xi_{3k_F}\ll\Delta^T_{\rm SE}$
($\xi_{3k_F}\gg\Delta^T_{\rm SE}$), the strength for the SO order parameter is enhanced
(suppressed) due to the enhanced effective attractive potential $V^{\rm at}_{\bf
k}\propto|\omega^{pl}_{\bf k}|^2\propto{n}$ (suppressed
pairing potential as mentioned in Sec.~\ref{SE-order}), and then a density peak of the strength
for the SO order parameter is expected. This can
    also be understood from Eq.~(\ref{SOO}). By
approximately taking the Coulomb interaction $V^0_{\bf k-k'}\omega^{pl}_{\bf
  k-k'}$ as a delta-function, it is found $|\Delta_{\rm
  SO}({\bf k}=3k_F{\bf e_x},\omega)|\propto\sqrt{nE_F}E_F/(8E^2_F+|\Delta^T_{\rm
SE}|^2)^2$, showing peak behavior in the density dependence. 

Moreover, from Fig.~\ref{figyw14}, it is found that at fixed density, the
strength of the SO order parameter $|\Delta_{\rm
  SO}({\bf k}=3k_F{\bf e_x},\omega)|$ increases with the frequency, in
consistent with the analytic results in Eq.~(\ref{SOO}).

\begin{figure}[htb]
  {\includegraphics[width=9cm]{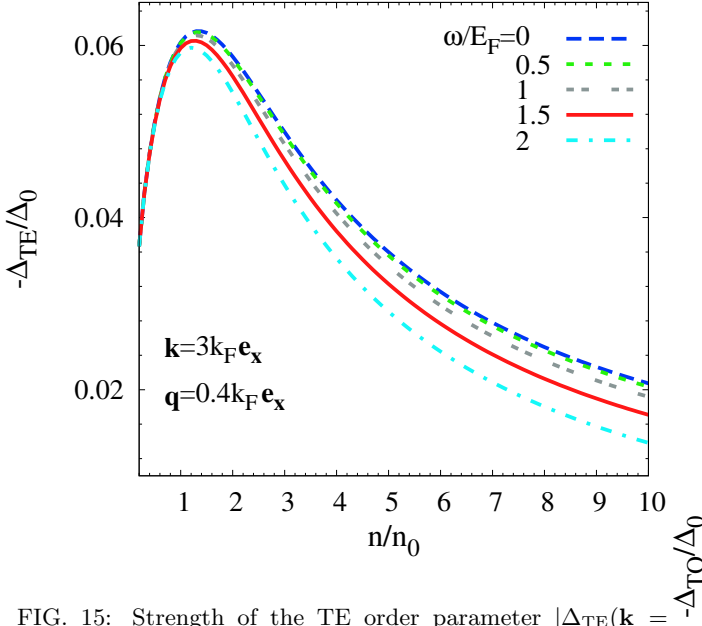}}
\caption{Strength of the TE order parameter $|\Delta_{\rm
  TE}({\bf k}=3k_F{\bf e_x},\omega)|=-\Delta_{\rm
  TE}({\bf k}=3k_F{\bf e_x},\omega)$ versus density $n$ at different
frequencies. ${\bf q}=0.4k_F{\bf e_x}$.}   
\label{figyw15}
\end{figure}

\subsection{Density dependence of the TE order parameter} 
 
We next address the density dependence of the
TE order parameter.  The strengths of
 TE order parameters $|\Delta_{\rm
  TE}({\bf k}=3k_F{\bf e_x},\omega)|$ versus the density $n$ are
plotted in Fig.~\ref{figyw15} at different
frequencies. As shown from the figure, with the increase of the density,  
$|\Delta_{\rm TE}({\bf
  k}=3k_F{\bf e_x},\omega)|$ first increases when $n<1.2n_0$ then decreases
after $n>1.2n_0$, leading to
a peak around $1.2n_0$, similar to the results of the SO order parameter. 

The density dependence of $|\Delta_{\rm
  TE}({\bf k}=3k_F{\bf e_x},\omega)|$ can be understood from the momentum-magnitude 
dependence, since the increase of the density leads to the increase of
$3k_F$. Then, the density dependence shows similar behavior of the momentum-magnitude 
dependence, and the density peak (shown in Fig.~\ref{figyw15}) corresponds to
the momentum-magnitude peak above (shown in Fig.~\ref{figyw7}).

In addition, we also find that at fixed density, the strength of the
TE order parameter $-\Delta_{\rm TE}({\bf k}=3k_F{\bf e_x},\omega)$ decreases 
with the increases of the frequency, as shown in Fig.~\ref{figyw15}, in good
agreement with Eq.~(\ref{TE-re}).

\begin{figure}[htb]
  {\includegraphics[width=9cm]{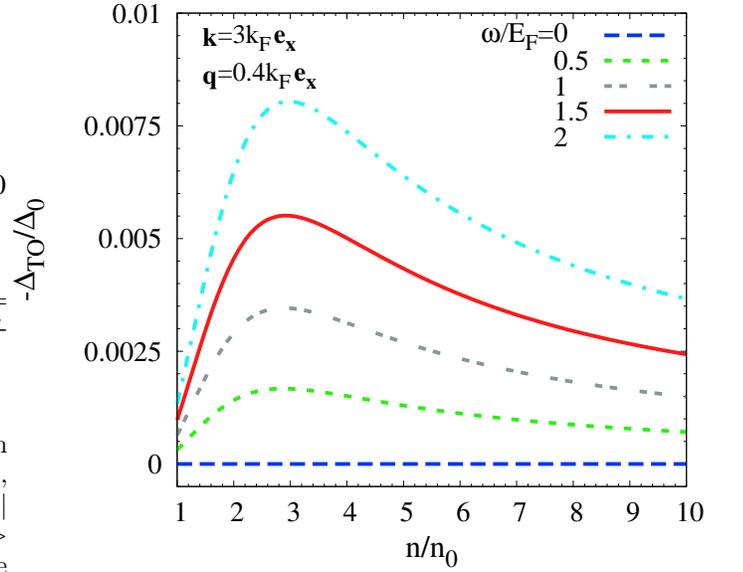}}
\caption{Strength of the TO order parameter $|\Delta_{\rm TO}({\bf
    k}=3k_F{\bf e_x},\omega)|=-\Delta_{\rm TO}({\bf
    k}=3k_F{\bf e_x},\omega)$ versus density $n$ at different
frequencies when ${\bf 
    q}=0.4k_F{\bf e_x}$.}   
\label{figyw16}
\end{figure}

\subsection{Density dependence of the TO order 
  parameter} 

We then address the density dependence of the
TO order parameter.  The strengths of
TO order parameter are plotted against the density $n$ in Fig.~\ref{figyw16} at different
frequencies. As shown from the figure, with the increase of the density,  
a peak is observed in the density dependence of $\Delta_{\rm TO}({\bf
    k}=3k_F{\bf e_x},\omega)$, similar to the previous results of the SO and TE order
  parameters. 

With the similar analysis for the density dependence of the SO and TE order
parameter, the strength for the TO one is enhanced 
(suppressed) by the increase of the density at small (large) density 
due to the enhanced effective attractive potential $V^{\rm at}_{\bf
k}$ and SOC $h_{3k_F{\bf e_x}}$ (suppressed
pairing potential as mentioned in Sec.~\ref{SE-order}), leading to a peak observed. 

Moreover, from Fig.~\ref{figyw16}, it is found that at fixed density, the strength of the
TO order parameter $|\Delta_{\rm TO}|$ increases with the frequency, in
consistent with the analytic results in Eq.~(\ref{TOE}).

\end{appendix}

\end{document}